\documentclass[superscriptaddress, twocolumn, amsmath, amssymb, aps, pra]{revtex4-2} 
\usepackage[english,french]{babel} % or [french,english] if you want French default
\usepackage[T1]{fontenc} % optional with XeLaTeX, can remove too
\usepackage{graphicx} 
\usepackage{caption} 
\usepackage{parskip} 
\usepackage{ulem}
\usepackage{cases}
\usepackage{multirow}
\usepackage{physics}
\usepackage{xcolor}
\usepackage{booktabs}
\usepackage[T1]{fontenc} 
\usepackage{subcaption}
\usepackage[justification=raggedright,singlelinecheck=false]{caption}
\usepackage{ulem} 
\usepackage{parskip} 
\usepackage[dvipsnames]{xcolor}
\usepackage{ragged2e}

\begin{document}

\title{Two-photon-assisted collisions in ultracold gases of polar molecules II : \\
Optical shielding of ultracold polar molecular collisions} 

\author{Gohar Hovhannesyan}
\affiliation{Universit\'e Paris-Saclay, CNRS, Laboratoire Aim\'e Cotton, 91400 Orsay, France}
\author{Charbel Karam}
\affiliation{Universit\'e Bourgogne Europe, CNRS, Laboratoire Interdisciplinaire Carnot de Bourgogne ICB UMR 6303, 21000 Dijon, France}
\author{Romain Vexiau}
\affiliation{Universit\'e Paris-Saclay, CNRS, Laboratoire Aim\'e Cotton, 91400 Orsay, France}
\author{Leon Karpa}
\affiliation{Institut für Quantenoptik, Leibniz Universität Hannover, 30167 Hannover, Germany}
\author{Maxence Lepers}
\affiliation{Universit\'e Bourgogne Europe, CNRS, Laboratoire Interdisciplinaire Carnot de Bourgogne ICB UMR 6303, 21000 Dijon, France}
\author{Nadia Bouloufa-Maafa}
\affiliation{Universit\'e Paris-Saclay, CNRS, Laboratoire Aim\'e Cotton, 91400 Orsay, France}
\author{Olivier Dulieu}
\affiliation{Universit\'e Paris-Saclay, CNRS, Laboratoire Aim\'e Cotton, 91400 Orsay, France}

%\date{} % Leave empty to omit a date

\begin{abstract}

We theoretically investigate the collisions between ultracold polar molecules in the presence of two lasers ensuring a Raman resonant transition on individual molecules to suppress photon scattering, taking the example of bosonic $^{23}$Na$^{39}$K molecules. By varying laser detunings and intensities, we enable a repulsive long-range interaction potential between molecules. After solving a set of coupled Schr\"odinger equations with the Hamiltonian written in the basis of laser-dressed states of the molecule pair at infinite distance, we identify quasi-resonant conditions under which elastic collisions are favored over inelastic and reactive ones, by a factor of about 2, thus demonstrating a promising pathway for efficient two-photon optical shielding of ultracold molecular collisions. The results are analyzed in terms of scattering length of the colliding laser-dressed molecules, which exhibit prominent resonances assigned to the interaction of the entrance channel with other specific channels, consistent with the existence of a quasi-bound level of the long-range molecular pair induced by the lasers. 
\end{abstract}

\maketitle
\section{Introduction}

Being a quantum system itself, the laser is a universal tool for controlling ensembles of quantum particles such as neutral atoms and molecules, ions, and Rydberg atoms. With such a statement, we paraphrase the title and content of the review article written by the Nobel Prize laureate in Physics Serge Haroche: ``Laser, offspring and powerful enabler of quantum science'' \cite{haroche2025}. Among the numerous achievements of the last decades are undoubtedly laser cooling and trapping of neutral atoms, which ultimately lead to the observation of quantum degeneracy in dilute gases, as well as exquisite control of ensembles of Rydberg atoms with giant anisotropic interactions. In addition, with their rich internal structure, including an intrinsic electric and/or magnetic dipole moment, neutral “polar'' molecules soon appeared as promising candidates to further extend these developments \cite{doyle2004,carr2009,dulieu2009,quemener2012,moses2017,bohn2017, shi2025bose} in fundamental physics \cite{karman2024} as well as in the context of future quantum technologies \cite{cornish2024,demille2024}. However, experimental manipulation of ultracold molecule samples towards quantum degeneracy is hindered by the rapid loss of particles because of their short-range interactions. On the one hand, this has led to spectacular results in ultracold quantum chemistry, where reactions can be controlled by quantum effects \cite{liu2022bimolecular}. On the other hand, these losses occur even when molecules are prepared in their absolute ground state, where there is no energetically allowed exit channel \cite{ospelkaus2010a, takekoshi2014, molony2014, gregory2019, park2015}. Current models tentatively explain these losses by the formation of a transient four-body complex with high densities of internal states, often referred to as "sticky" collisions \cite{mayle2013,bause2023}.

A general strategy has been devised to prevent such losses based on the engineering of intermolecular interactions at large distances with electromagnetic fields, endowing them with a repulsive nature, and thus tuning the intermolecular scattering length. This has first been proposed \cite{avdeenkov2006a,wang2005,gonzalez-martinez2017} and observed \cite{li2021,mukherjee2023} using strong static electric fields. Another option relies on microwave (MW) transitions that couple two rotational levels of the electronic ground state of the molecule \cite{lassabliere2018,karman2018}. Such a microwave shielding (MW-S) has been beautifully observed in experiments, leading to the creation of a degenerate quantum gas of MW-dressed molecules \cite{anderegg2021,schindewolf2022,bigagli2023}, but at the expense of a somewhat tedious control of the MW field. 

While the above approaches have demonstrated their efficiency to stabilize ultracold molecular samples, the use of lasers for such quantum engineering emerges as an untapped yet promising path that deserves to be explored. One-photon optical shielding (1-OS) \cite{xie2020,xie2022} uses a laser with a frequency detuned to the blue of a molecular rovibronic transition that couples colliding ground-state molecules to a repulsive state involving a ground-state molecule and an excited-state one. The proof of principle of such a one-photon optical shielding (1-OS) has been demonstrated for inelastic collisions between ultracold alkali-metal atoms \cite{bali1994, marcassa1994, suominen1996a, hoffmann1996}. However, in the case of molecules, 1-OS is limited by spontaneous photon scattering, which leads to undesirable heating \cite{xie2020}. To prevent such a drawback, we proposed \cite{karam2023} a two-photon optical shielding (2-OS) involving a two-photon $\Lambda$-type transition between rovibronic levels of a single molecule, operating in the far red-detuned electromagnetically induced transparency (EIT) regime so that the role of the upper level of the $\Lambda$ scheme can be adiabatically eliminated. Although the possibility of influencing the course of molecular collisions has been assessed, shielding efficiency has not yet been demonstrated \cite{karam2025}. 

In this work, we extend the study of \cite{karam2025} (hereafter referred to as paper I) to 2-OS in the weakly red-detuned or blue-detuned EIT regime in order to explicitly involve the upper level of the $\Lambda$ scheme in the molecular dynamics as an additional control knob. The paper is organized as follows: in Section \ref{sec:5level} we recall the main features of 2-OS of two interacting molecules colliding in the presence of two lasers, modeled as a five-level scheme. The Hamiltonian of two interacting ultracold molecules exposed to two coherent lasers is presented in Section \ref{sec:theory}. The resulting potential energy curves (PECs) describing the long-range interactions between the molecules are displayed in Section \ref{sec:DPEC}. The rate coefficients for elastic, inelastic and reactive collisions between the molecules are extracted in Section \ref{sec:dynamics} from a full quantum dynamical calculation, leading to the definition of the efficiency parameter for 2-OS. Section \ref{sec:discussion} discusses the prospects for improving the shielding efficiency and for experimental implementation.

\section{The dynamical five-level scheme for 2-OS}
\label{sec:5level}

\begin{figure}[!h]
   \centering 
    \includegraphics[scale=0.45]{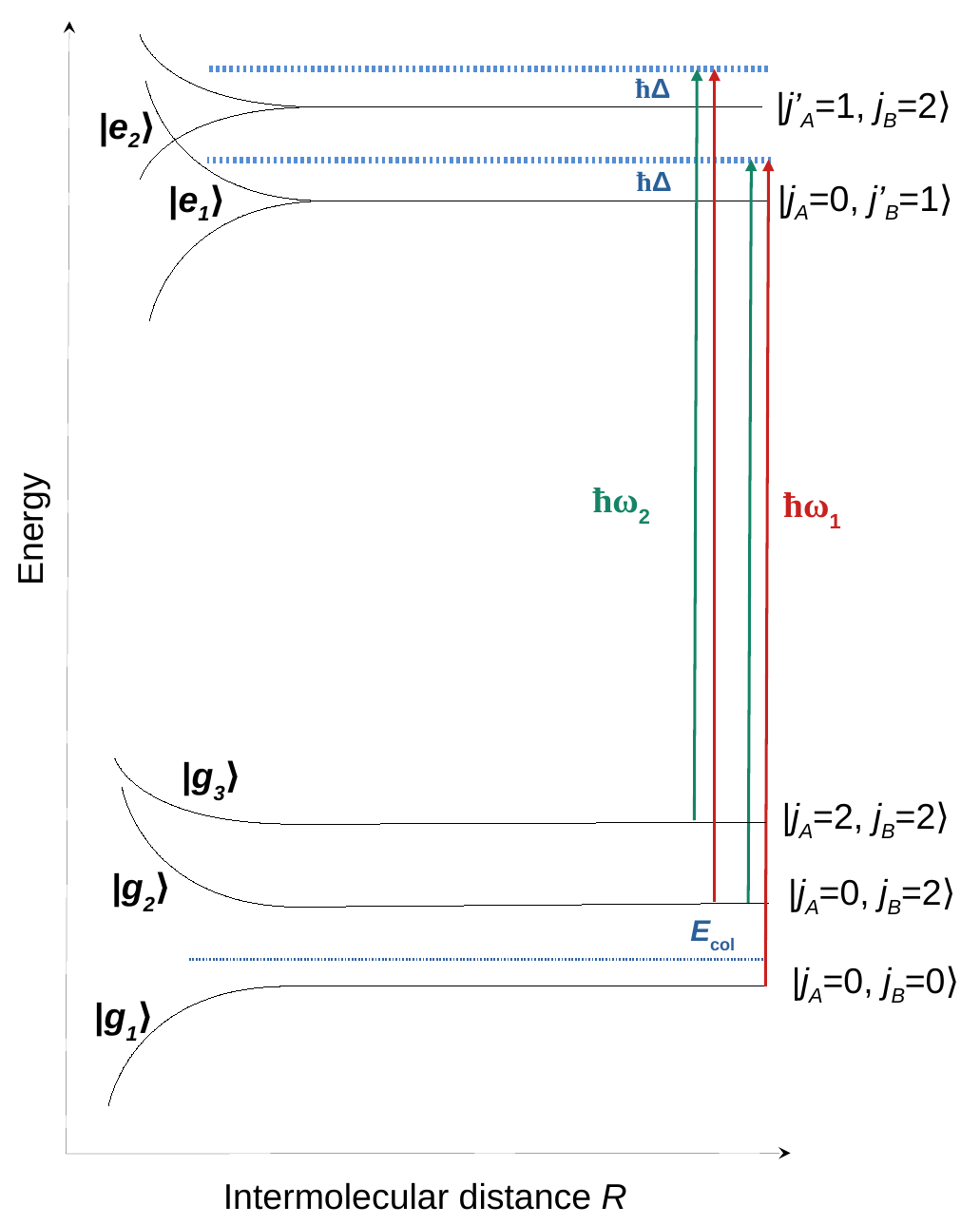}
    \caption{
    \parbox{\linewidth}{ 
            \justifying
            Schematic view of the five-level system relevant for 2-OS in the case of blue-detuned lasers. Energies are not to scale. The two photons with energies $\hbar \omega_1$ and $\hbar \omega_2$ (upper arrows) are detuned by a small detuning $\Delta$ with respect to the molecular transitions $\ket{j_i=0} \rightarrow \ket{j'_i=1}$ and $\ket{j_i=2} \rightarrow \ket{j'_i=1}$. ($i=A,B$) The sketched long-range potential energy curves (PECs) describe the interaction between two polar molecules colliding with the energy $E_{\mathrm{col}}$ in the space-fixed frame, characterized by their rotational quantum numbers $j_A$ and $j_B$ (see text for more details about the chosen system). This defines the five relevant coupled states $\ket{g_1}$, $\ket{g_2}$, $\ket{g_3}$, $\ket{e_1}$, $\ket{e_2}$, labeled at infinite distances by the rotational quantum numbers of the molecules. At all distances, the attractive entrance channel $\ket{g_1}$ is coupled to the repulsive channel $\ket{g_2}$ via the excited channel $\ket{e_1}$ by the two photons. The same coupling scheme occurs between $\ket{g_2}$ and $\ket{g_3}$ via $\ket{e_2}$.} }
    \label{fig:2_OS_scheme}
\end{figure}

The central concept of the 2-OS scheme is to engineer repulsive interactions between colliding ground-state ultracold molecules while minimizing photon scattering and spontaneous emission. Individual molecules are exposed to two coherent lasers $L_1$ and $L_2$ with circular frequencies $\omega_1$ and $\omega_2$, and Rabi frequencies $\Omega_1$ and $\Omega_2$ under a static EIT condition applied on a $\Lambda$-type scheme of three levels $|j\rangle$, $|j''\rangle$, $|j'\rangle$, with energies $E_{j}$, $E_{j''}$, $E_{j'}$. Assuming linearly-polarized lasers, the two former levels are rotational levels of the vibronic ground state of the molecule (typically, $j=0$ and $j''=2$), while the latter level is a rotational level of an electronically-excited state of the molecule (typically, $j'=1$). The detuning $\Delta$ of the lasers is considered positive (negative) for red (blue) detunings, namely: $\hbar \omega_1=E_{j'}-E_{j}-\hbar \Delta$, and $\hbar \omega_2=E_{j'}-E_{j''}-\hbar \Delta$ (see Fig. \ref{fig:2_OS_scheme}). At large distances, molecules interact with each other via dipole-dipole interaction (DDI) and van der Waals interaction (VWI), illustrated by the schematic PECs in Fig. \ref{fig:2_OS_scheme}. The entrance scattering channel for two molecules A and B, $\ket{g_1}$ (identical to $\ket{j_A=0;j_B=0}$ when $R \rightarrow \infty$) is attractive, and it is coupled by the two photons to the $\ket{g_2}$ channel ($\ket{g_2}_{R \rightarrow \infty} = \ket{j_A=0;j_B=2}$) chosen to be repulsive \cite{karam2023} to induce optical shielding. The excited state channel $\ket{e_1}$ ($\ket{e_1}_{R \rightarrow \infty} = \ket{j_A=0; j'_B=1})$ belongs to the excited manifold of the bimolecular complex and could be attractive or repulsive \cite{karam2023,karam2024}. When $R \rightarrow \infty$, the system $\{\ket{g_1},\ket{e_1},\ket{g_2} \}$ is composed of three levels of the molecule $B$, while the molecule $A$ is a spectator.  A similar system of three coupled states is composed of $\ket{g_2}$, $\ket{g_3}$ ($\ket{g_3}_{R \rightarrow \infty} = \ket{j_A=2;j_B=2}$), and $\ket{e_2}$ ($\ket{e_2}_{R \rightarrow \infty} = \ket{j'_A=1; j_B=2}$, where the molecule $A$ is now a spectator. Then, the three channels $\ket{g_1}$, $\ket{g_2}$, $\ket{g_3}$ are equidistant at infinity, separated by an energy equal to 6 times the rotational constant of the individual molecule. 

In paper I, we established that at infinite distance, the bimolecular system is represented by two degenerate static three-level systems, which give rise to a ''dynamical five-level system'' differing from the conventional five-level-static scheme \cite{li2011b} in that the resonance condition varies with the scattering coordinate $R$. We presented the complete formalism describing two-photon-assisted collisions between ultracold polar molecules, assuming that the collisional dynamics essentially takes place at a large intermolecular distance $R$. In paper I, we used $\Delta$ a magnitude comparable to or larger than Rabi frequencies $\Omega_1$ and $\Omega_2$ and positive, so that the $\ket{e_1}$ and $\ket{e_2}$ channels were eliminated adiabatically from the dynamics. Here, we consider $\Delta \ll \Omega_1, \Omega_2$ small, positive or negative, so that the $\ket{e_1}$ and $\ket{e_2}$ channels are explicitly involved in the dynamics. 

Following \cite{karam2023, xie2020} and paper I, we developed our study with bosonic molecules $^{23}$Na$^{39}$K prepared in the lowest rovibrational level ($v_X = 0$, $j_X = 0$) of their electronic ground state, $X^1\Sigma^+$ (hereafter referred to as $X$). The hyperfine structure of the molecule is neglected \cite{aldegunde2017}. For the two-photon transition, we consider the lowest electronic excited state of $^{23}$Na$^{39}$K $b^3\Pi_0$ (denoted $b$ hereafter). The $b$ state is weakly mixed with the $A^1\Sigma^+$ state (hereafter referred to as the $A$ state) via spin-orbit coupling, resulting in two electronic states of $\Omega = 0^+$ symmetry, where $\Omega$ is the projection of the total electronic angular momentum onto the internuclear axis. As a result, the $X \leftrightarrow b$ transition becomes dipole-allowed through the $A$ component of the mixed $0^+$ state, leading to a transition electric dipole moment (TEDM) of 0.0456~a.u. (or 0.116~Debye) for the $v_X = 0 \leftrightarrow v_b = 0$ transition \cite{xie2020}. The rotational constants of these levels are $B_X=0.0950$ cm$^{-1}$ and $B_b=0.0951$ cm$^{-1}$, respectively \cite{xie2020}.

\section{Hamiltonian of two interacting ultracold molecules exposed to two coherent lasers} 
\label{sec:theory}

The Hamiltonian $\hat{H}_{\text{int}}(R)$ for the interaction of two molecules ($i = A, B$) colliding under the influence of two laser fields ($\alpha = 1, 2$) is 
\begin{align} \label{eq:H_int}
\hat{H}_{\text{int}}(R) & = \sum_{i=A}^B \hat{H}_0(i) + \hat{V}_{\text{rot}}(R) + \hat{V}_{\text{int}}(R) \nonumber \\
& + \sum_{\alpha=1}^2 \hat{H}_{L}(\alpha)
\ + \sum_{\alpha=1}^2 \sum_{i=A}^B \hat{H}^{(\alpha)}_{\text{ac}}(i).
\end{align}
The Hamiltonian $\hat{H}_0(i)$ for the molecule $i$ results in a scalar term $E_0(i) = E_e(i) + E_{\text{vib}}(i) + B(i) \hat{j}_i^2$ (in atomic units) which includes the electronic energy $E_e(i)$, the vibrational energy $E_{\text{vib}}(i)$ and the rotational energy $B(i) \hat{j}_i^2$, where $\hat{j}_i$ is the angular momentum operator for the molecule $i$ with quantum number $j_i$, projection $m_i$ and associated parity $p_i=+/-$, and  $B(i)$ the rotational constant of the levels $v_X=0$ or $v_b=0$. The term $\displaystyle \hat{V}_{\text{rot}}(R)=\hbar^2 \hat{L}^2/(2 \mu R^2)$ represents the relative rotation of molecules with quantum numbers $\ell$ and $m_{\ell}$ and $\hat{V}_{\text{int}}(R)$ is the molecule-molecule interaction potential. The term $\hat{H}_{L}(\alpha)$ holds for the energy associated with the laser field $\alpha = L_1, L_2$, and $\hat{H}^{(\alpha)}_{\text{ac}}(i)$ the interaction between the molecule and the laser field $\alpha$.

We consider the uncoupled basis set $|\beta_A j_A m_A p_A, \beta_B j_B m_B p_B, \ell m_\ell\rangle$ for the angular momenta of each molecule and of their mutual rotation (which can be removed at infinite distance) completed by the Fock states denoting the number of photons in each laser field $\ket{N_1,N_2}$, written as tensor products: 
\begin{equation}
    \ket{n}= |\beta_A j_A m_A p_A, \beta_B j_B m_B p_B, \ell m_\ell\rangle \otimes \ket{N_1, N_2}
    \label{eq:barestates}
\end{equation}
where $\beta_i$ ($i=A, B$) includes the quantum numbers characterizing the electronic state of each molecule,that is, $\beta_i \equiv X, b$. The basis set is symmetrized with respect to the exchange of the two molecules $A$ and $B$. At infinite distance, the five states of Fig.\ref{fig:2_OS_scheme} describing the two molecules are written as:  
\begin{equation}
\begin{cases}
    \ket{g_1} = \ket{ X,0,0,+, X,0,0,+} \otimes \ket{N_1, N_2}_{+}, \\
    \ket{g_2} = \ket{ X,0,0,+, X,2,0,+} \otimes \ket{N_1-1, N_2 + 1}_{+}, \\
    \ket{g_3} = \ket{ X,2,0,+, X,2,0,+} \otimes \ket{N_1-2, N_2 + 2}_{+}, \\
    \ket{e_1} = \ket{ X,0,0,+, b,1,0,-} \otimes \ket{N_1-1, N_2}_{-},  \\
    \ket{e_2} = \ket{ X,2,0,+, b,1,0,-} \otimes \ket{N_1-2, N_2 + 1}_{-}.
  \end{cases}
  \label{eq:bas5st}
\end{equation}
where we assume a linear polarization of the lasers, ignore the quantum numbers $\ell$ and $m_{\ell}$, and display the global parity index $+/-$ for each vector. When the molecules interact, these five vectors are interacting with many other basis vectors such that the induced Hilbert space can be described as five ensembles (or Floquet blocks) of basis vectors identified by a well-defined number of photon: 
\begin{equation}
\begin{aligned}
&\{\mathcal{G}_1\} = \{|X j_A m_A p_A, X j_B m_B p_B, \ell m_{\ell}\rangle \otimes |N_1, N_2 \rangle_+\}, \\
&\{\mathcal{G}_2\} = \{|X j_A m_A p_A, X j_B m_B p_B, \ell m_{\ell}\rangle \otimes | N_1 - 1, N_2 + 1 \rangle_+\}, \\
&\{\mathcal{G}_3\} = \{|X j_A m_A p_A, X j_B m_B p_B, \ell m_{\ell}\rangle \otimes | N_1 - 2, N_2 + 2 \rangle_+\}, \\
&\{\mathcal{E}_1\} = \{|b j_A m_A p_A, b j_B m_B p_B, \ell m_{\ell}\rangle \otimes | N_1 - 1, N_2 \rangle_-\}, \\
&\{\mathcal{E}_2\} = \{|b j_A m_A p_A, b j_B m_B p_B, \ell m_{\ell}\rangle \otimes | N_1 - 2, N_2 + 1 \rangle_-\}.
\end{aligned}
\label{eq:subspaces_uncoupled}
\end{equation}
where each subspace $\{\mathcal{G}_k\} (k=1\cdots 3)$ and $\{\mathcal{E}_k\} (k=1,2)$ contains a set of rotational states $j_i, m_i$ and partial waves $\ell, m_{\ell}$ that produce a fixed value of the projection $M=m_a+m_B+m_{\ell} \equiv 0$ of the total angular momentum $\hat{J}=\hat{j}_A+\hat{j}_B+\hat{L}$. The state $\ket{g_1}$ being the entrance channel in our study (Fig. \ref{fig:2_OS_scheme}), the subspaces $\{\mathcal{G}_k\}$ involve states of total parity ``$+$'' imposing that $j_A+j_B+\ell$ is even, while the subspaces $\{\mathcal{E}_k\}$ are composed of states with parity ``$-$'' (and thus odd values of $j_A+j_B+\ell$) because of (E1) selection rules. In addition, permutation symmetry demands even values of $\ell$. 

The matrix of the Hamiltonian $\hat{H}_{\text{int}}$ of Eq.\ref{eq:H_int} is structured in five blocks in this basis as
\begin{widetext}
\begin{equation} \label{eq:H_total}
\mathbf{H}_{\text{int}} = \begin{pmatrix}
\mathbf{H_g}(R) & 0 & 0 & \mathbf{V_1} & 0 \\
 & \mathbf{H_g}(R) - \hbar \delta\omega & 0 & \mathbf{V_2} & \mathbf{V_1} \\
 & & \mathbf{H_g}(R) - 2\hbar \delta\omega & 0 & \mathbf{V_2} \\
 & & & \mathbf{H_e}(R) - \hbar \omega_{1} & 0 \\
 & & & & \mathbf{H_e}(R) + \hbar(\omega_{2} - 2\omega_{1}) 
\end{pmatrix}, 
\end{equation} 
\end{widetext}
where $\hbar \delta\omega = \hbar(\omega_{1} - \omega_{2}) \equiv 6B_X$ (Fig. \ref{fig:2_OS_scheme}). The submatrices $\mathbf{H_g}$ and $\mathbf{H_e}$ contain the matrix elements of $\hat{H}_0$, $\hat{V}_{\text{rot}}$ and $\hat{V}_{\text{int}}$ for two ground-state molecules ($X(v_X = 0)$) and for one ground-state molecule and an excited one ($b(v_b = 0)$). The five blocks correspond to the three $\{\mathcal{G}_k\}$ and two $\{\mathcal{E}_k\}$ subspaces, and include energy shifts induced by $\hat{H}_{L}(\alpha)$ due to the variation of the number of photons in the Fock states with respect to $\{\mathcal{G}_1\}$. The terms $V_1$ and $V_2$ represent the light couplings of the colliding pair with the first and second lasers generated by $\hat{H}^{(\alpha)}_{\text{ac}}$.

The configuration space spanned by the ensembles $\{\mathcal{G}_k\}$ and $\{\mathcal{E}_k\}$ of Eq. \ref{eq:subspaces_uncoupled} is built by considering values of the angular quantum numbers varying from 0 to $j_{1\text{max}}=4$, $j_{2\text{max}}=4$, $\ell_{\text{max}}=4$ (and thus $J_{\text{max}}=12$), following the convergence tests performed in \cite{karam2024,karam2025}. Considering that the initial state of the system is the s-wave ($\ell=0, m_{\ell}=0$) between two molecules in their absolute ground state ($j_A=j_B=0$, $m_A=m_B=0$), we chose $M=m_A+m_B+m_{\ell}=0$ for the ensemble $\{\mathcal{G}_1\}$. By considering two linearly polarized laser fields, under the (E1) selection rules, we take $M=0$ for all the other ensembles. We therefore consider all possible values of $m_A$, $m_B$ and $m_{\ell}$ provided $m_A + m_B + m_{\ell} = 0$. A total of 1528 basis vectors is generated in the uncoupled basis set: 170 for each set ${\mathcal{G}_k}$ with even inversion, permutation, and reflection symmetry, and 509 for each set ${\mathcal{E}_k}$ with odd inversion but even permutation and reflection symmetry. 

\section{Laser-dressed bimolecular potential energy curves} 
\label{sec:DPEC}

As stated in the introduction, we are interested in the collision of two ultracold molecules exposed to two coherent lasers tuned to an appropriate Raman resonance of a single molecule, in the regime of weak blue detuning $\Delta \ll \Omega_1, \Omega_2$, equivalent to a strong coupling regime. By minimizing $\Delta$, and thus enhancing the laser-induced couplings between $\{\mathcal{G}_k\}$ and $\{\mathcal{E}_k\}$, we explore the possibility for the excited state manifold to yield additional flexibility for improved shielding efficiency. 

Following paper I, we first diagonalize the Hamiltonian of Eq.\ref{eq:H_total} to obtain dressed adiabatic PECs at large distances, typically $R>1000$~a.u.. Therefore, when $R \rightarrow \infty$, the individual molecules are dressed by the lasers, \textit{i.e.} the quantum state $ \ket{\tilde{n}}_{\infty}$ of the molecular pair at infinity is a linear combination of bare states $\ket{n}$, 
\begin{equation}
    \ket{\tilde{n}}_{\infty} = \sum_n c^{\infty}_{\tilde{n},n} \ket{n}, 
    \label{eq:dressed}
\end{equation}
where $({c^{\infty}_{\tilde{n},n}})^2$ determines the weight of the bare state $\ket{n}$. When the two lasers are tuned to a Raman resonance, the initial state of the dressed molecular pair corresponds to two molecules in the resulting dark state. If the Hamiltonian of  Eq.\ref{eq:H_int} is diagonalized in the basis of the five levels of Eq.\ref{eq:bas5st}, this dark state is unshifted with respect to the energy of $\ket{g_1}$ taken as the origin of the energies. We note that the energy of the dark state resulting from the diagonalization of the matrix in Eq.\ref{eq:H_int} will be slightly shifted with respect to the origin as the two lasers interact with other energy levels in a strongly off-resonant way: $L_2$ (resp. $L_2$) is off-resonant with the transition $j=0 \rightarrow j'=1$ (resp. $j=2 \rightarrow j'=1$). Thus, the initial five-level system of Eq. \ref{eq:bas5st} interacts with 
$ \ket{ X,0,0,+, X,2,0,+} \otimes \ket{N_1, N_2}_{+}$, $\ket{ X,0,0,+, X,0,0,+} \otimes \ket{N_1-1, N_211}_{+}$, and $\ket{ X,0,0,+, b,3,0,-} \otimes \ket{N_1-2, N_2-1}_{-}$, located respectively at $\approx 17$~GHz, $\approx -17$~GHz, and $\approx 12$~GHz.

\begin{figure}[!h]
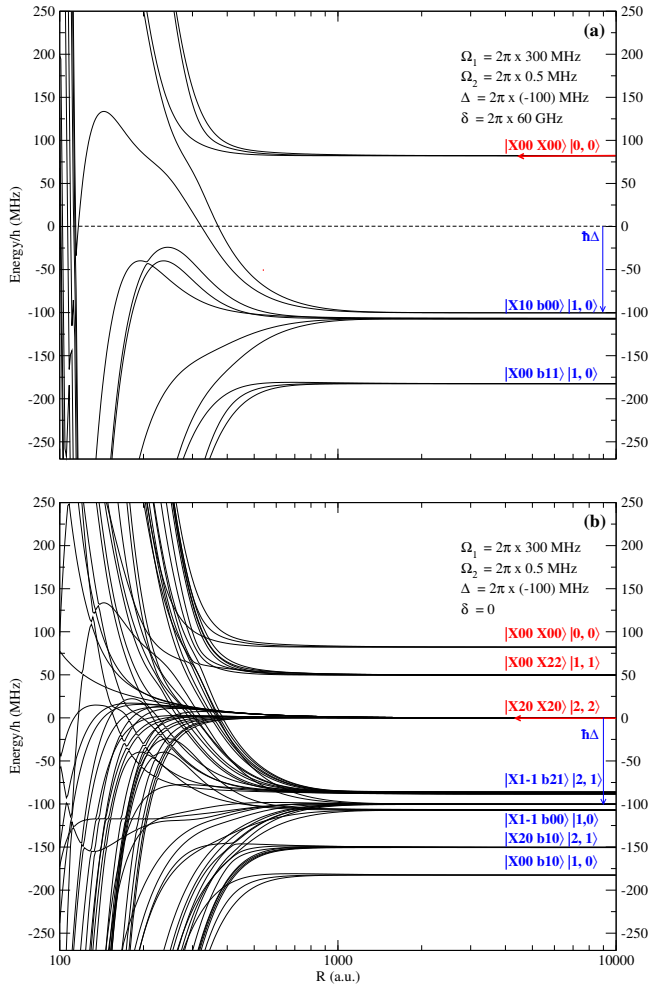

    \centering 
    \includegraphics[scale=0.32]{dressed_PECs_O_300_05_D_100_petit_d_60GHz_v5}
    \includegraphics[scale=0.32]{dressed_PECs_O_300_05_D_100_petit_d_0_v5}
    \caption{\parbox{\linewidth}{ 
            \justifying
            Adiabatic dressed PECs of two interacting NaK molecules prepared in the $\ket{j_X=0}$ level of the rovibronic ground state exposed to two lasers with $\Omega_1=2\pi \times 300$~MHz and $\Omega_2=2\pi \times 0.5$~MHz, and a blue detuning $\Delta=-2\pi \times 100$~MHz pictured by a downward blue arrow. (a) The laser $L_2$ is detuned by a large amount $\delta= 2\pi \times 60$~GHz from the Raman resonance. (b) The two lasers are tuned to the chosen Raman resonance ($\delta=0$, see text). The curves are labeled with the basis vector $\ket{n}$ that has the largest component in the expansion of Eq.\ref{eq:dressed} for $R \rightarrow \infty$. }}
    \label{fig:Dressed_pecs_non_eff}
\end{figure}

As we work with blue detunings $\Delta$, we first check if our model recovers the limiting case of 1-OS \cite{xie2020}, choosing a large value $\Delta=-2\pi \times 100$~MHz. This is obtained by setting $\Omega_2$ to a weak value and by detuning the laser $L_2$ by a large amount $\delta$ so that the Raman resonance condition is deliberately not fulfilled. The results are displayed in Fig. \ref{fig:Dressed_pecs_non_eff}a: it reflects a dynamics similar to that described in \cite{xie2020} \footnote{Note that in \cite{xie2020}, the Rabi frequency $\Omega_1$ was at least 3 times smaller, while only the $s$, $p$ and $d$ waves were involved in the scattering calculations.} with the entrance channel exhibiting a marked avoided crossing with three excited channels, inducing efficient shielding of the molecular collisions.

If the Raman resonance is set ($\delta=0$) with a laser $L_2$ with a weak Rabi frequency, the PECs become more complex (Fig. \ref{fig:Dressed_pecs_non_eff}b) as more eigenstates are close in energy at infinite distances. Moreover, the entrance channel is made up of two molecules in the EIT dark state, hence mostly in the $j=2$ rotational level. There are no large avoided crossings between the entrance channel and other repulsive PECs, reflecting the weak interactions induced by the laser light, as observed in paper I for red detunings. 

\begin{figure}[!ht]
    \centering
   \includegraphics[scale=0.4]{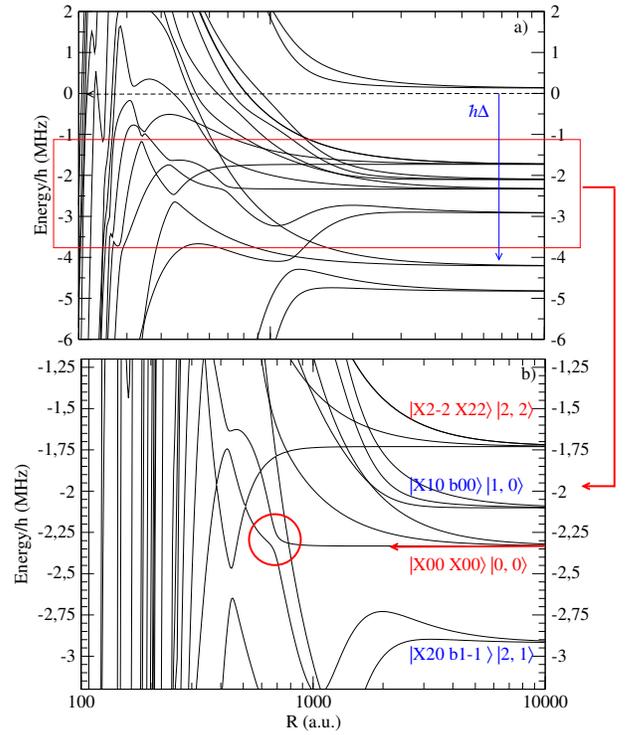}
    \caption{\parbox{\linewidth}{ 
            \justifying
            (a) Adiabatic dressed PECs of two interacting NaK molecules prepared in the $\ket{j_X=0}$ level of the rovibronic ground state exposed to two lasers with $\Omega_1=2\pi \times 456$~MHz and $\Omega_2=2\pi \times 831$~MHz, and a blue detuning $\Delta=-2\pi \times 4$~MHz (pictured by a downward blue arrow). (b) Enlarged view of the zone inside the red box in panel (a). The curves are labeled as in Fig. \ref{fig:Dressed_pecs_non_eff}. The red arrow shows the entrance channel for the collision. The choice of these specific values for the laser parameters is explained in Section \ref{sec:dynamics}.}}
    \label{fig:dressed_blue_D_4_O_456_831_ID}
\end{figure}

The adiabatic dressed PECs for $\Delta \ll \Omega_1, \Omega_2$ are shown in Fig. \ref{fig:dressed_blue_D_4_O_456_831_ID}. The chosen values of the laser parameters will become clear in the next section. For such small $\Delta$ values, the PECs are quite close to each other, generating numerous avoided crossings and quite complex behavior. It appears that in this representation the entrance channel is indeed coupled to repulsive channels via marked avoided crossings, but of small amplitude. The consequences for the dynamics of 2-OS are discussed in the next section. 

\section{Quasi-resonant 2-OS in the regime of small detunings}
\label{sec:dynamics}

In order to obtain the scattering rates for the two colliding molecules at large distances, we solve the time-independent Schr\"odinger equation for the energy $E_{\mathrm{col}}$ with the interaction Hamiltonian matrix of Eq.\ref{eq:H_int} re-expressed in the dressed asymptotic basis of Eq.\ref{eq:dressed}. In this representation the interaction with light is diagonal, while the off-diagonal terms result from the dipole-dipole and van der Waals interactions. After extracting the $S$-matrix, we calculate the rate coefficients for elastic collisions $\beta_i^{\text{el}}$, for inelastic collisions $\beta_i^{\text{inel}}$, and for reactive collisions  $\beta_i^{\text{rea}}$ as follows: 
\begin{align}
  \beta_i^{\text{el}} (E_{\mathrm{col}}) &= g_i \frac{\pi}{\mu k_i} |1-S_{ii}(E_{\mathrm{col}})|^2
  \label{elastic rates} \\
  \beta_i^{\text{inel}} (E_{\mathrm{col}}) &= g_i \frac{\pi}{\mu k_i}
    \sum_{j \neq i} |S_{ij}(E_{\mathrm{col}})|^2 
  \label{inelastic rates} \\
  \beta_i^{\text{rea}} (E_{\mathrm{col}}) &= g_i \frac{\pi}{\mu k_i}
    \left( 1 - \sum_j |S_{ij}(E_{\mathrm{col}})|^2 \right)
  \label{reactive rates}
\end{align}
where the index $i$ stands for the entrance channel of the collision, namely the dark state of the bimolecular pair at infinity, and with  $g_i=2$ for identical colliding species and $g_i=1$ otherwise. The initial relative wave vector $k_i \equiv (2 \mu E_{\mathrm{col}})^{1/2}/\hbar$ of the molecule pair with reduced mass $\mu$ is taken for a temperature of $E_{\mathrm{col}}/k_B \equiv T=300$~nK representative of a typical ultracold molecule experiment. The shielding efficiency parameter is introduced: 
\begin{equation}
    \gamma_i (E_{\mathrm{col}})= \frac{\beta_i^{\text{el}}(E_{\mathrm{col}})}{\beta_i^{\text{inel}}(E_{\mathrm{col}}) + \beta_i^{\text{rea}}(E_{\mathrm{col}})}
\end{equation}
which reflects, if it is greater than 1, the dominant contribution of elastic collisions against   inelastic and reactive collisions. Inspired by \cite{xie2020} good shielding efficiency is achieved when $\gamma_i \approx 100$ or more. 

Following the extensive convergence states achieved in paper I, which allowed us to reasonably shorten the calculation for this study, we propagated the log-derivative of the wavefunction of the system from $R_{\mathrm{min}}=15$~a.u. to $R_{\mathrm{max}} = 10000$~a.u. using the method of Manolopoulos \cite{manolopoulos1986} using an adaptive grid step mapping the dressed PECs \cite{kokoouline1999, willner2004}. We impose an absorbing condition at $R_{\mathrm{min}}$ modeling the experimentally observed losses from \cite{Wang2015}. 

As exposed in paper I, particular care must be taken to identify the entrance channel. In a pure five-level system, the dark state of the bimolecular complex at infinity would be the one exhibiting no energy shift induced by the light interaction. In the present multichannel representation, the dark state is slightly shifted in energy by an amount $E_i$ due to the interaction with neighboring Fock blocks, resulting in a typical shift of a few MHz in the present calculations. Due to the complexity of the system exemplified by Fig. \ref{fig:dressed_blue_D_4_O_456_831_ID}, there are many channels with such a small energy. We identify the entrance channel as the one that contains a sizable component of the $\ket{g_1}$ state (Eq.\ref{eq:bas5st}). We checked that there is only one channel of that kind with an energy close to 0. Then the collisional channels are energetically open if their energy $E(\ket{\tilde{n}}_{\infty})$ is less than the total energy of the entrance channel $E_i^{\mathrm{tot}} = E_i + E_{\mathrm{col}}$. For example, in the conditions of Fig.\ref{fig:dressed_blue_D_4_O_456_831_ID}, the entrance channel is identified with the following weights on the basis vectors: 0.528 on $\ket{00, 00, \ell,m_\ell} \otimes \ket{0, 0}$, 0.397 on $\ket{20, 00, \ell,m_\ell} \otimes \ket{-1, +1}$, and 0.074 on $\ket{20, 20, \ell,m_\ell} \otimes \ket{-2, +2}$.

\begin{figure}[!h]
\centering
\includegraphics[width=0.94\linewidth]{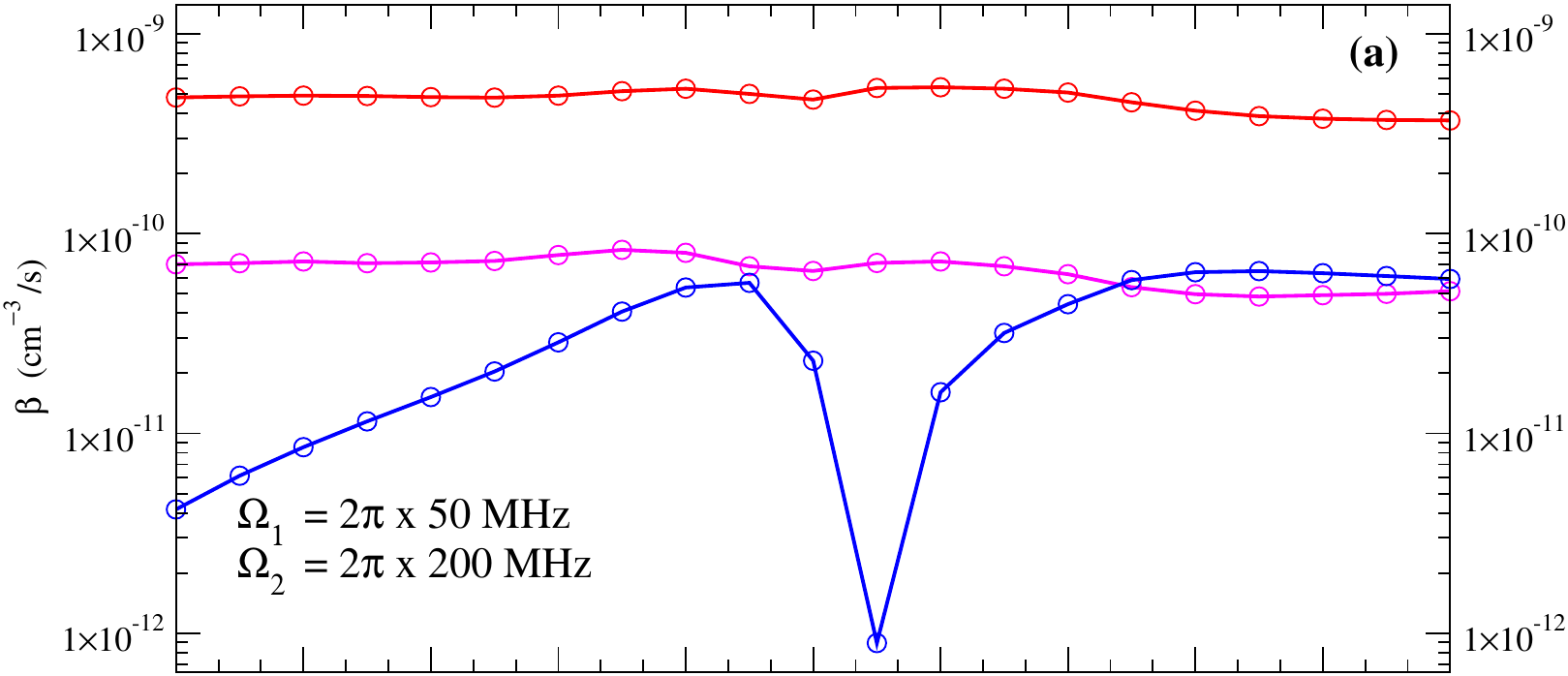}
\includegraphics[width=0.94\linewidth]{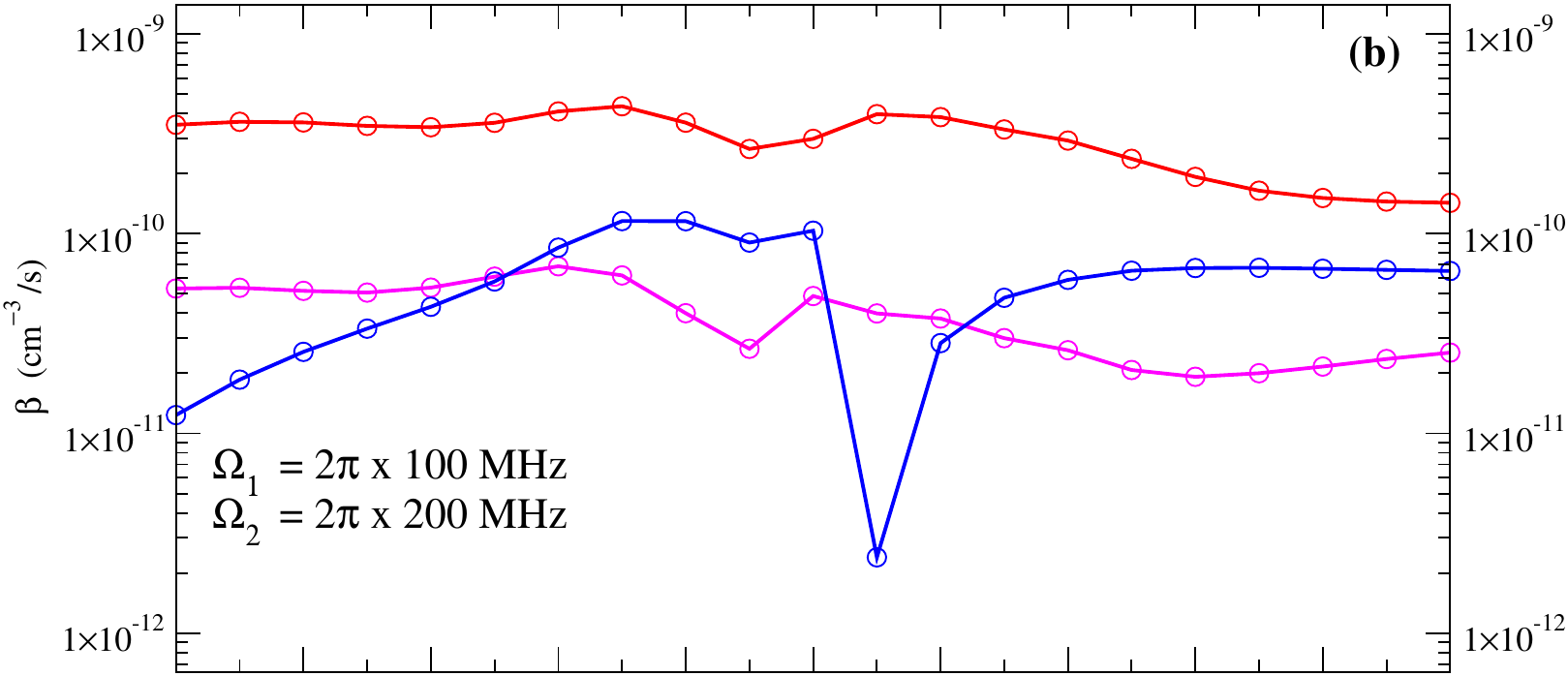}
\includegraphics[width=0.94\linewidth]{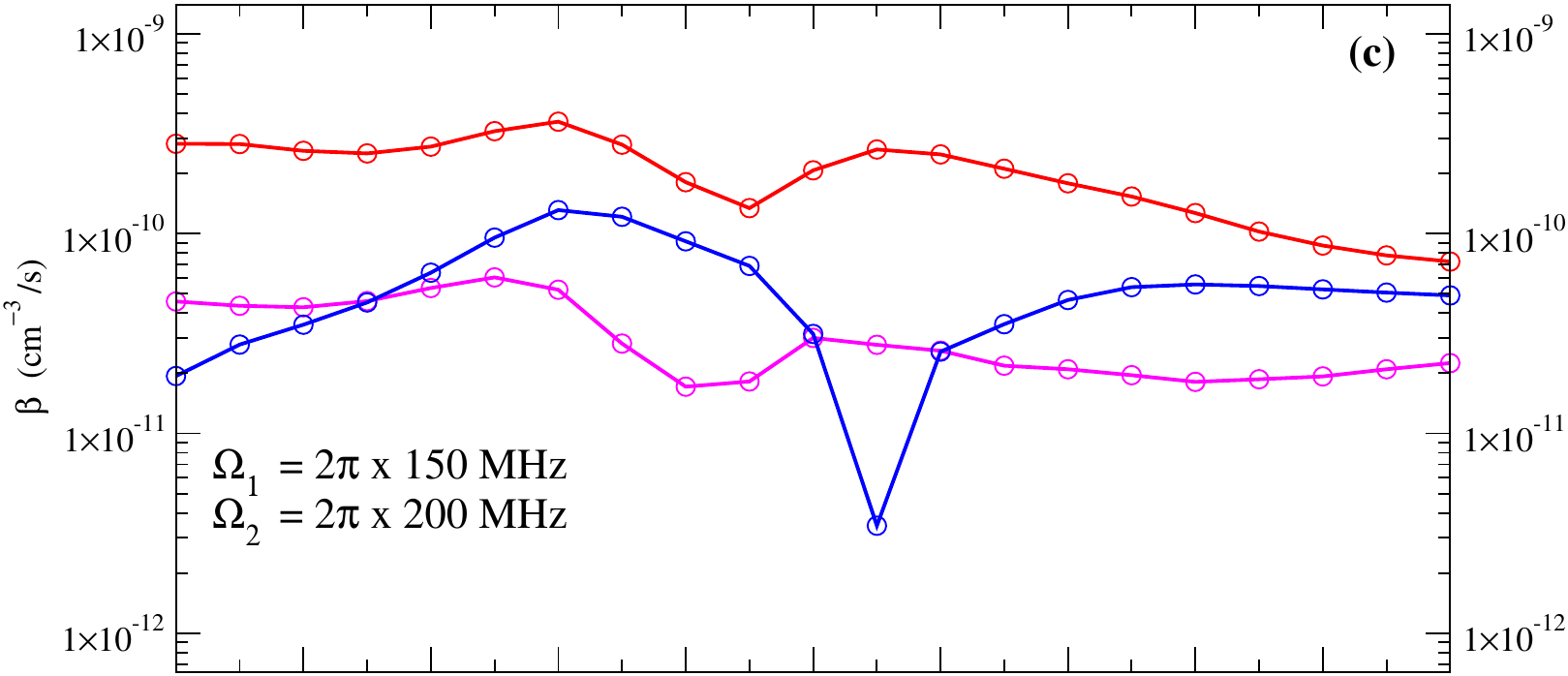}
\includegraphics[width=0.94\linewidth]{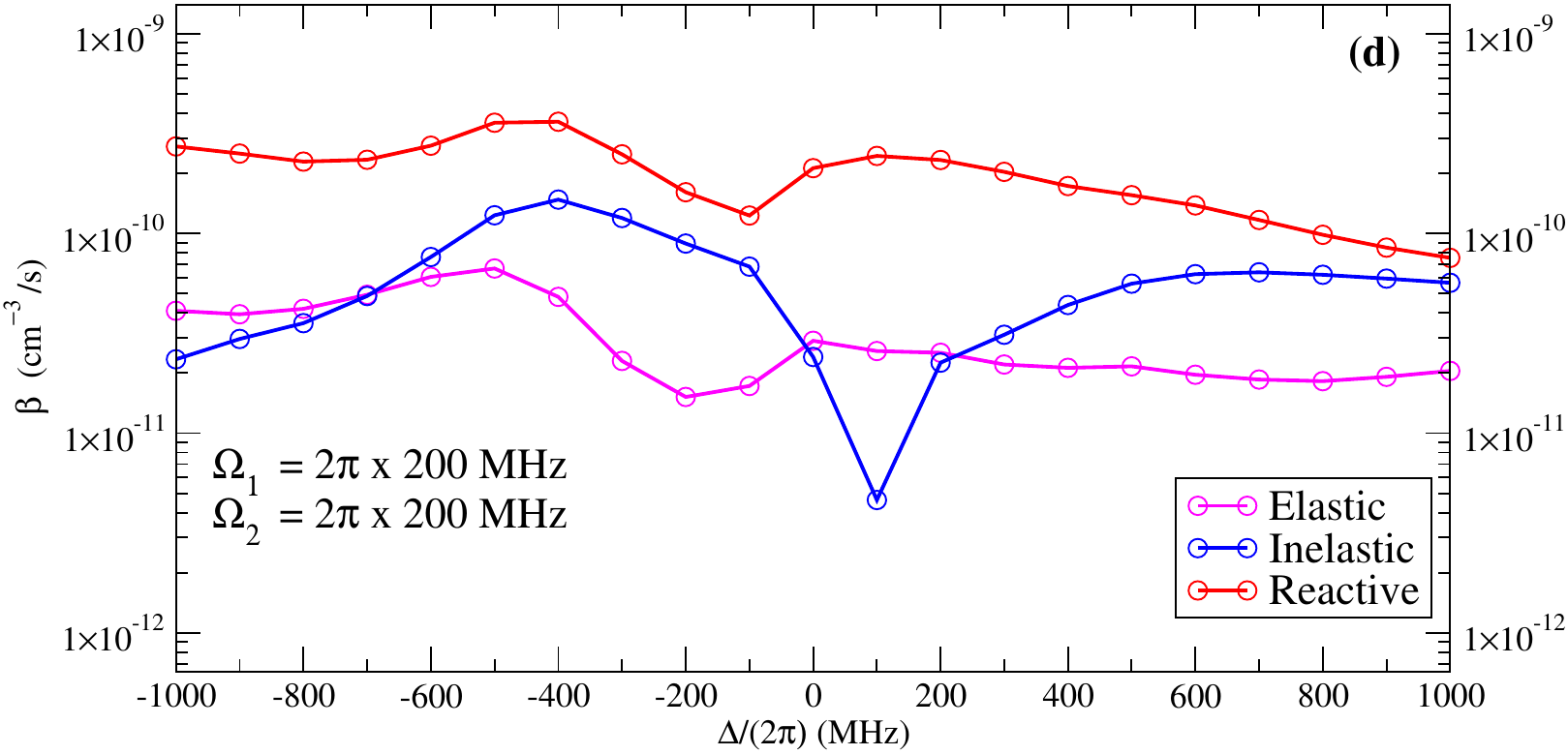}
\caption{\parbox{\linewidth}{\justifying
Collision rate coefficients $\beta_i^{\text{el}}$ (magenta), $\beta_i^{\text{inel}}$ (blue),
and $\beta_i^{\text{rea}}$ (red) (all labeled as $\beta$ on the vertical
axis, for simplicity) at $E_{\mathrm{col}} = k_B \times 300$~nK as functions
of $\Delta$ for $\Omega_2=2 \pi \times 200$~MHz and (a) $\Omega_1=2 \pi \times 50$~MHz, (b) $2 \pi \times 100$~MHz, (c) $2 \pi \times 150$~MHz, (d) $2 \pi \times 200$~MHz.}}
\label{fig:rates_scan_detuning}
\end{figure}

We first investigate the situation corresponding to Fig.\ref{fig:Dressed_pecs_non_eff} with $\Delta$ not negligible with respect to $\Omega_1$ and $\Omega_2$. Figure \ref{fig:rates_scan_detuning} displays the various rate coefficients for a fixed value of $\Omega_2=2\pi \times 200$~MHz, $\Omega_1$ varying between $2\pi \times 50$~MHz and $2\pi \times 200$~MHz, and for $\Delta/(2\pi)$ varying from +1~GHz (red detuning) down to -1~GHz (blue detuning). The general trend of the rate coefficients is the same as discussed in paper I for red detunings: the reactive rate systematically exceeds the elastic rate for all detunings, at least with the chosen step for $\Delta$. This is not surprising, as the avoided crossings present in Fig.\ref{fig:Dressed_pecs_non_eff} are very small, that is, the PECs are separated by an energy much lower than $\hbar \times 0.1$~MHz, invisible at the scale of the figure. However, it seems that the hierarchy between the various rates now depends on $\Delta$ for blue detunings, which is strikingly in contrast to the monotonic behavior already observed with red detunings. Moreover, the inelastic rate coefficient exhibits a pronounced minimum around $\Delta=0$ that motivates the next step of our study. 

We therefore investigated the dynamics of the system for blue detunings close to 0, choosing a small step size for the laser parameters, to identify if there exist zones where $\gamma_i>1$. Figure \ref{fig:dressed_blue_D_4_O_456_831_ID} is somewhat encouraging in this respect, as it exhibits an example of an avoided crossing which seems to be more marked than those in Fig. \ref{fig:Dressed_pecs_non_eff}. However, it is likely that because of the complexity of the PECs, the dynamics will be determined by the interaction between many channels. A global repulsive behavior of the entrance channel is visible, which can be assigned to couplings with numerous low-lying molecular PECs.

\begin{figure}[t]
\centering

\includegraphics[width=0.92\linewidth]{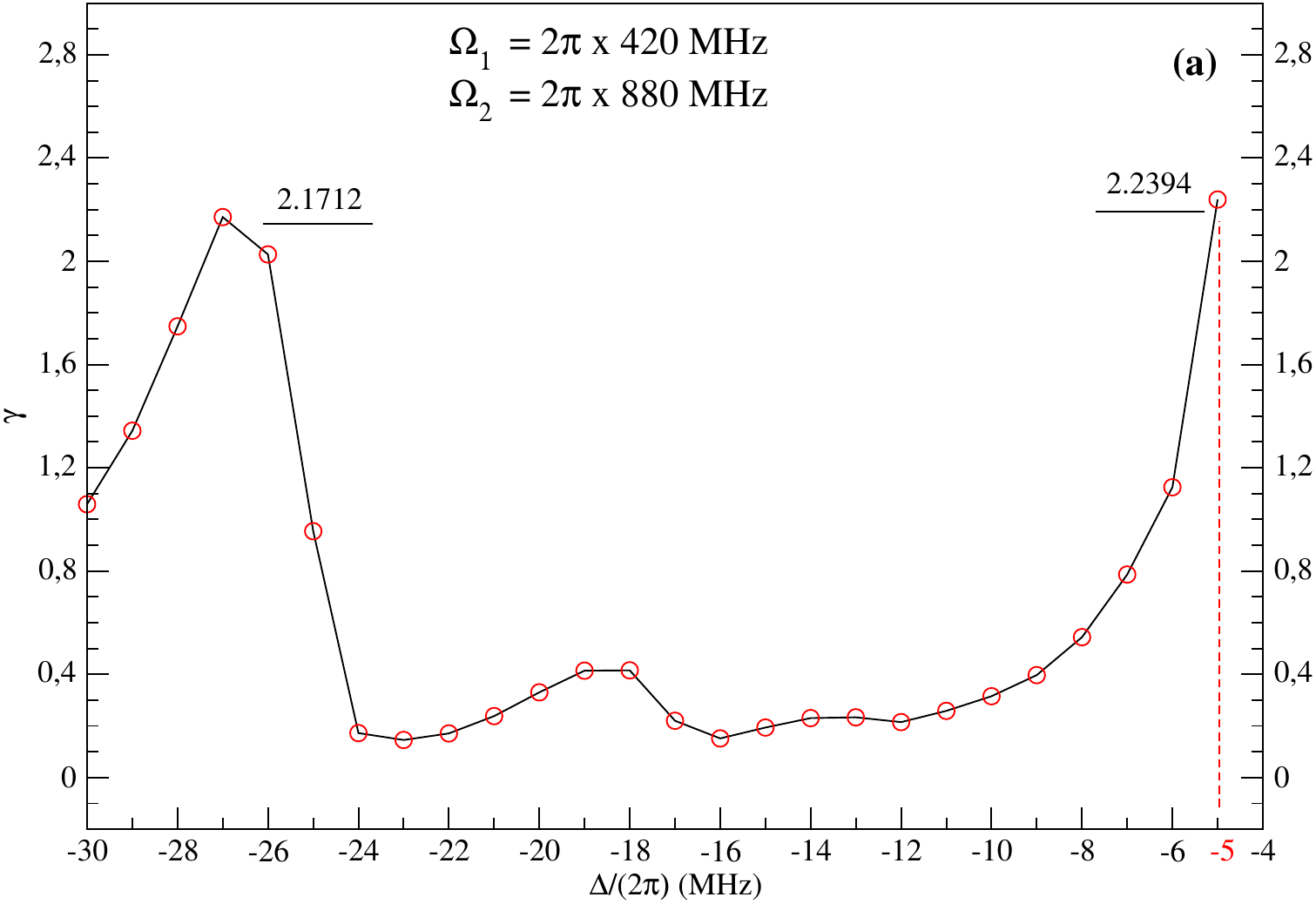}\\
\includegraphics[width=0.98\linewidth]{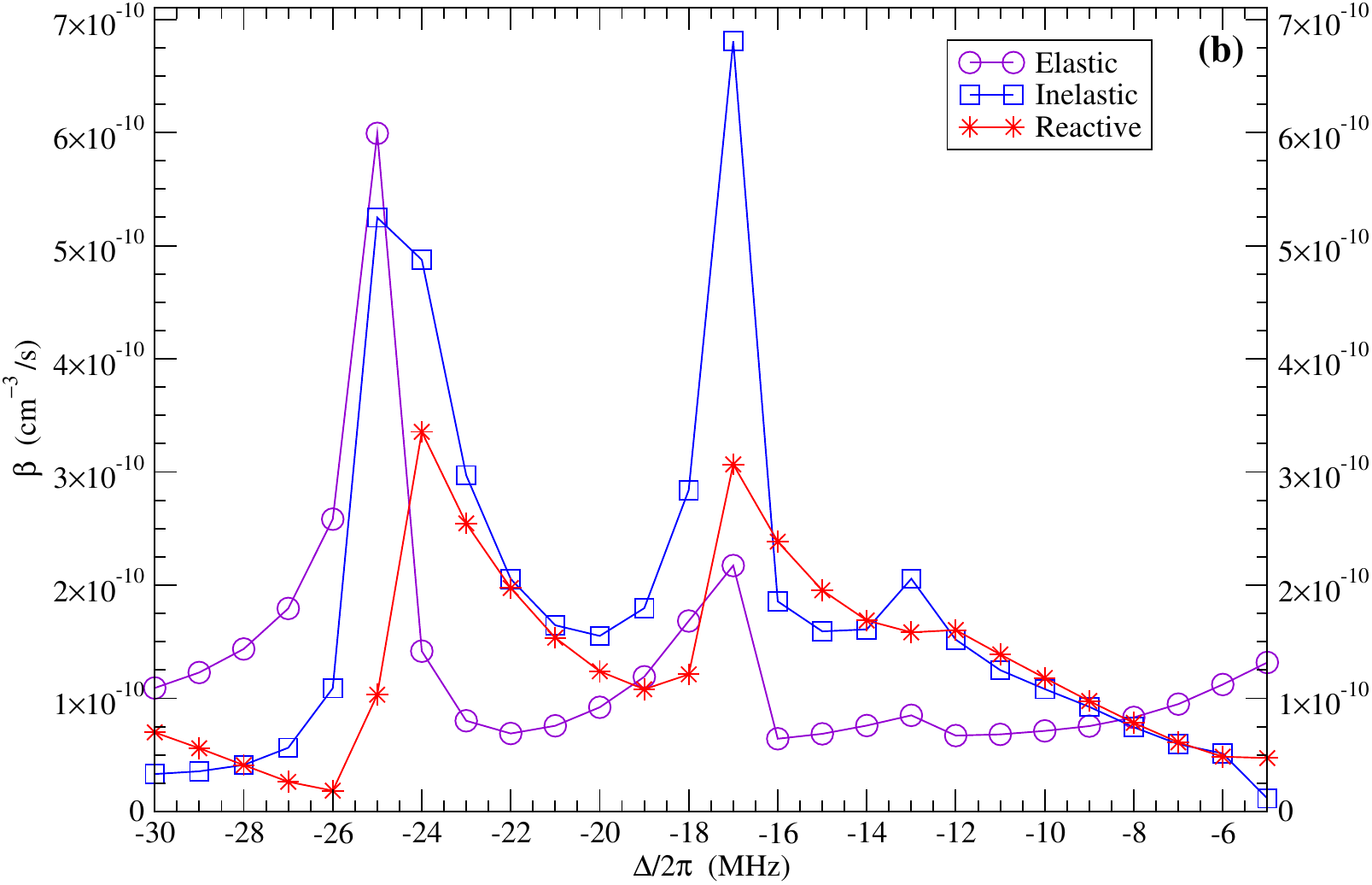}\\
\includegraphics[width=0.92\linewidth]{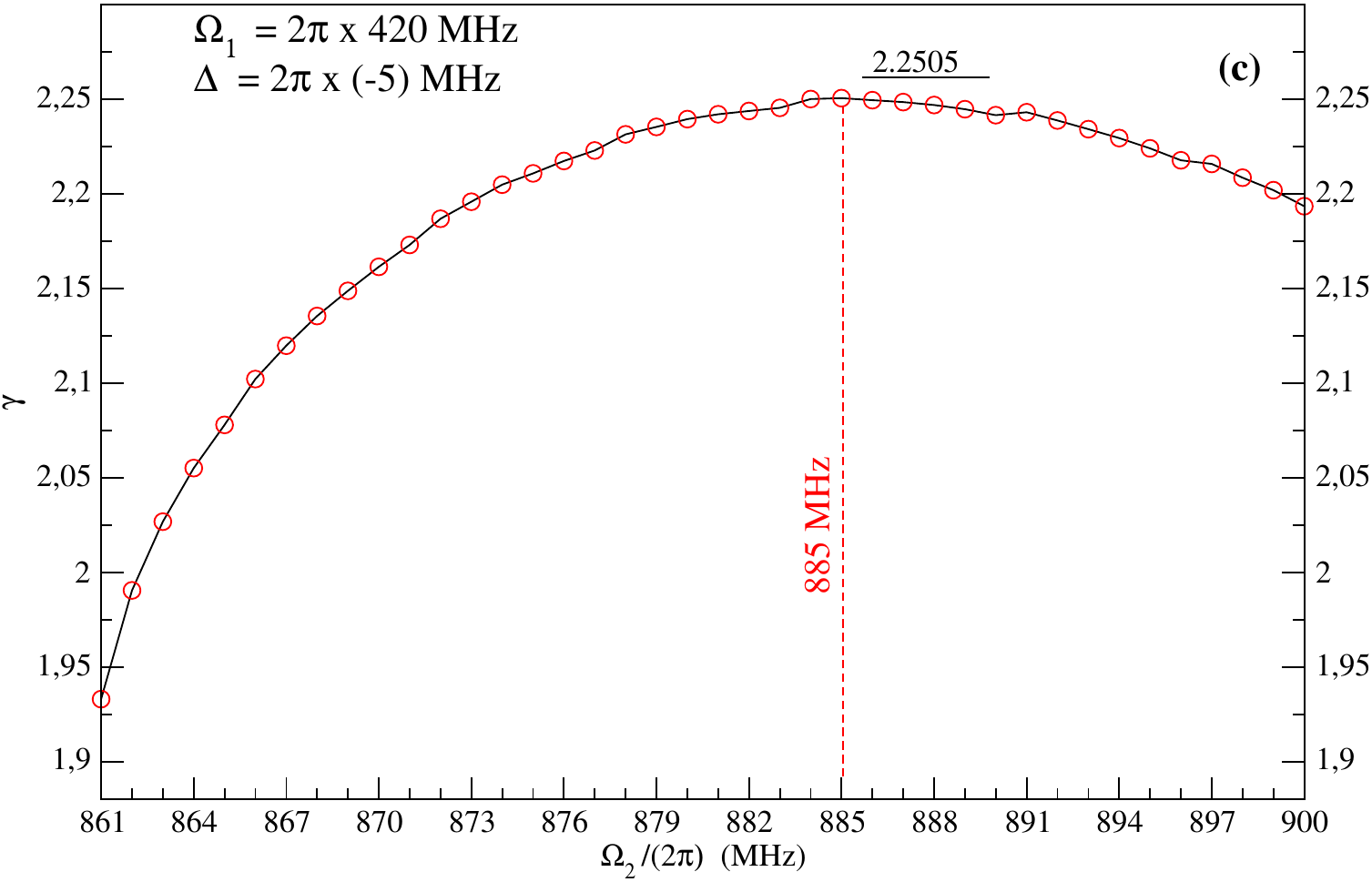}

\caption{\parbox{\linewidth}{\justifying
(a) Variation of the 2-OS efficiency parameter $\gamma_i$ as a function of $\Delta$ at an energy $E_{\mathrm{col}} = k_B \times 300$~nK, for the Rabi frequencies $\Omega_1/(2\pi)=420$~MHz and $\Omega_1/(2\pi)=880$~MHz which generates two peaks at two values of $\Delta$ where elastic collisions dominate. (b) Variation of the corresponding rate coefficients $\beta_i^{\text{el}}$ (magenta line), $\beta_i^{\text{inel}}$ (blue line), $\beta_i^{\text{rea}}$ (orange line) (all labeled as $\beta$ on the vertical axis, for simplicity). (c) Variation of $\gamma_i$ with $\Omega_2$ for $\Omega_1/(2\pi)=420$~MHz and $\Delta/(2\pi)=-5$~MHz.}}
\label{fig:gamma_D_loop_O_420_880}
\end{figure}

We display in Fig.\ref{fig:gamma_D_loop_O_420_880} a typical example of the variation of the rate coefficients (panel (b)) and the resulting $\gamma_i$ (panel (a)) over a restricted range of low $\Delta$ values. From our numerical survey of the parameter space ($\Omega_1$,$\Omega_2$), we identified couples of values producing peaks demonstrating that the elastic rate coefficient dominates the two other rate coefficients for well-defined values of $\Delta$. The values of $\gamma_i \approx 2$ is not yet large enough to implement an efficient 2-OS, but they convey the idea that we have to search for a quasi-resonant dynamics for the elastic collisions with respect to the detuning. Panel (b) suggests a typical width of about $\Delta/(2\pi) \approx 4$~MHz for this feature. In contrast, panel (c) of the same figure shows that the variation of the magnitude of such a peak in $\gamma_i$ with $\Omega_2$ is low, which might be of great interest in view of an experimental investigation.

%The model is tested on different parametric values of $\Omega_1$, $\Omega_2$ and $\Delta$ to find the region where shielding becomes effective.

\begin{figure*}[t]
    \centering
    \includegraphics[scale=0.42]{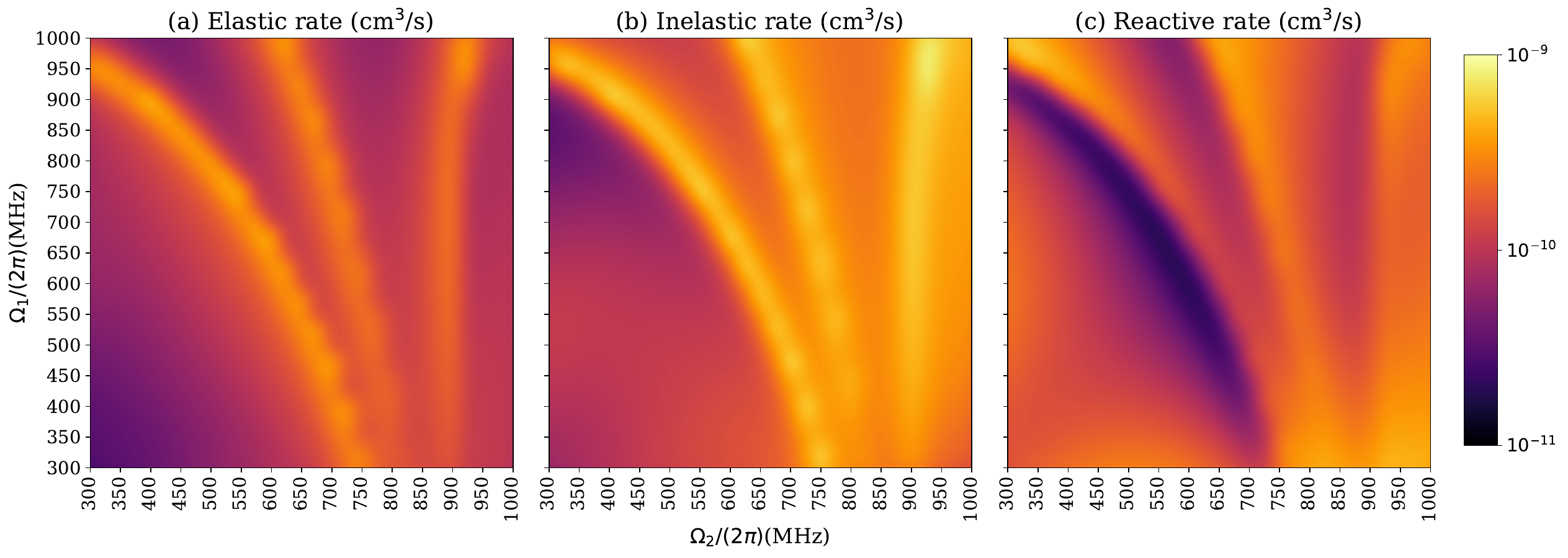}
    \caption{\parbox{\linewidth}{ 
            \justifying
            Color map in $\Omega_1/(2\pi)$ and $\Omega_2/(2\pi)$ with a step size of 25~MHz for collision rate coefficients $\beta_i^{\text{el}}$ (a), $\beta_i^{\text{inel}}$ (b), $\beta_i^{\text{rea}}$ (c) at an energy $E_{\mathrm{col}} = k_B \times 300$~nK and for a fixed detuning $\Delta/2\pi = -25$~MHz. The colored logarithmic scale extends from $10^{-11}$ to $10^{-9}$~$\text{cm}^3/\text{s}$.}}
    \label{fig:rates_color_map_D_25_step_50_25}
\end{figure*}

Table \ref{tab:data} reveals that such prominent features in $\gamma_i$ occur for several combinations of $\Omega_1$ and $\Omega_2$ where the selected combinations were chosen on the criterion of the highest values of $\gamma$ on a grid for $\Omega_1$ and $\Omega_2$ and for two representative values of $\Delta/(2\pi)=-25$~MHz and -5~MHz. We see that the dynamics is not identical for the two detunings: for the former value, $\beta_i^{\text{el}}$ largely exceeds $\beta_i^{\text{rea}}$ by a factor of about 10 and $\beta_i^{\text{inel}}$ only by a factor of about 2, while this pattern is reversed for the latter value. Therefore, the peak values of $\gamma_i$ are limited by the subtle balance between the inelastic and reactive dynamics, enforcing the quasi-resonant character of 2-OS. 

\begin{table}[ht!]
\centering
\begin{tabular}{cccccc}
\toprule
\textbf{$\Omega_1/(2\pi)$} & $\Omega_2/(2\pi)$ & $\gamma_i$& $\beta_i^{\text{el}}/\beta_i^{\text{inel}}$ & $\beta_i^{\text{el}}/\beta_i^{\text{rea}}$ \\  
\toprule 
\multicolumn{5}{c}{$\Delta/(2\pi)=-25$~MHz} \\ \hline
300 & 940   & 1.60 & 1.94 & 9.16  \\  
350 & 900   & 1.856 & 2.37 & 8.54  \\ 
360 & 900   & 1.82 & 2.18 & 11.12 \\  
380 & 880   & 1.91 & 2.34 & 10.35 \\
400 & 850   & 1.92 & 2.61 & 7.30  \\ 
420 & 860   & 1.70 & 1.94 & 13.60 \\ 
440 & 800   & 1.86 & 2.61 & 6.44  \\  
440 & 820   & 1.96 & 2.42 & 10.37 \\ 
440 & 840   & 1.72 & 1.96 & 14.08 \\ 
450 & 800   & 1.94 & 2.51 & 8.52  \\  
480 & 780   & 1.89 & 2.24 & 12.17 \\  
480 & 800   & 1.63 & 1.83 & 14.63 \\ 
500 & 750   & 1.87 & 2.23 & 11.54 \\  
500 & 760   & 1.82 & 2.11 & 13.40 \\ 
520 & 740   & 1.72 & 1.94 & 14.60 \\  
540 & 700   & 1.74 & 2.00 & 13.22 \\  
550 & 700   & 1.61 & 1.80 & 15.22   \\  
560 & 680   & 1.61 & 1.81 & 14.85 \\  
600 & 600   & 1.49 & 1.68 & 13.21  \\  
650 & 550   & 1.01 & 1.10 & 12.48   \\
\bottomrule
\multicolumn{5}{c}{$\Delta/(2\pi)=-5$~MHz} \\ \hline
300 & 880  & 1.65 & 7.48 & 2.11 \\
320 & 880  & 1.80 & 8.71 & 2.27 \\
340 & 880  & 1.95 & 9.86 & 2.43 \\
360 & 860  & 2.09 & 8.50 & 2.77 \\
360 & 880  & 2.07 & 10.73 & 2.57 \\
380 & 860  & 2.15 & 8.75 & 2.85 \\
380 & 880  & 2.17 & 11.31 & 2.69 \\
400 & 860  & 2.15 & 8.63 & 2.86 \\
400 & 880  & 2.24 & 11.59 & 2.77 \\
400 & 900  & 2.12 & 12.53 & 2.56 \\
420 & 880  & 2.24 & 11.38 & 2.79 \\
420 & 900  & 2.19 & 12.83 & 2.64 \\
440 & 900  & 2.19 & 12.82 & 2.64 \\
460 & 940  & 1.86 & 8.98 & 2.35 \\
480 & 880  & 1.92 & 5.60 & 2.92 \\
500 & 880  & 2.00 & 7.28 & 2.76 \\
500 & 900  & 2.08 & 7.93 & 2.82 \\
520 & 900  & 2.09 & 9.73 & 2.66 \\
540 & 920  & 1.98 & 9.12 & 2.54 \\
560 & 940  & 1.63 & 6.18 & 2.21 \\
\bottomrule
\end{tabular}
\caption{\parbox{\linewidth}{ 
            \justifying
            Selected sets of $\Omega_1/(2\pi)$ and $\Omega_2/(2\pi)$ values for the two detunings $\Delta = -25$~MHz and $\Delta = -5$~MHz, for which the effective parameter $\gamma_i>1$. The last two columns display the ratios $\beta_i^{\text{el}}/\beta_i^{\text{inel}}$ and $\beta_i^{\text{el}}/\beta_i^{\text{rea}}$. }}
\label{tab:data}
\end{table}

\begin{figure}[ht]
    \centering
   \includegraphics[scale=0.36]{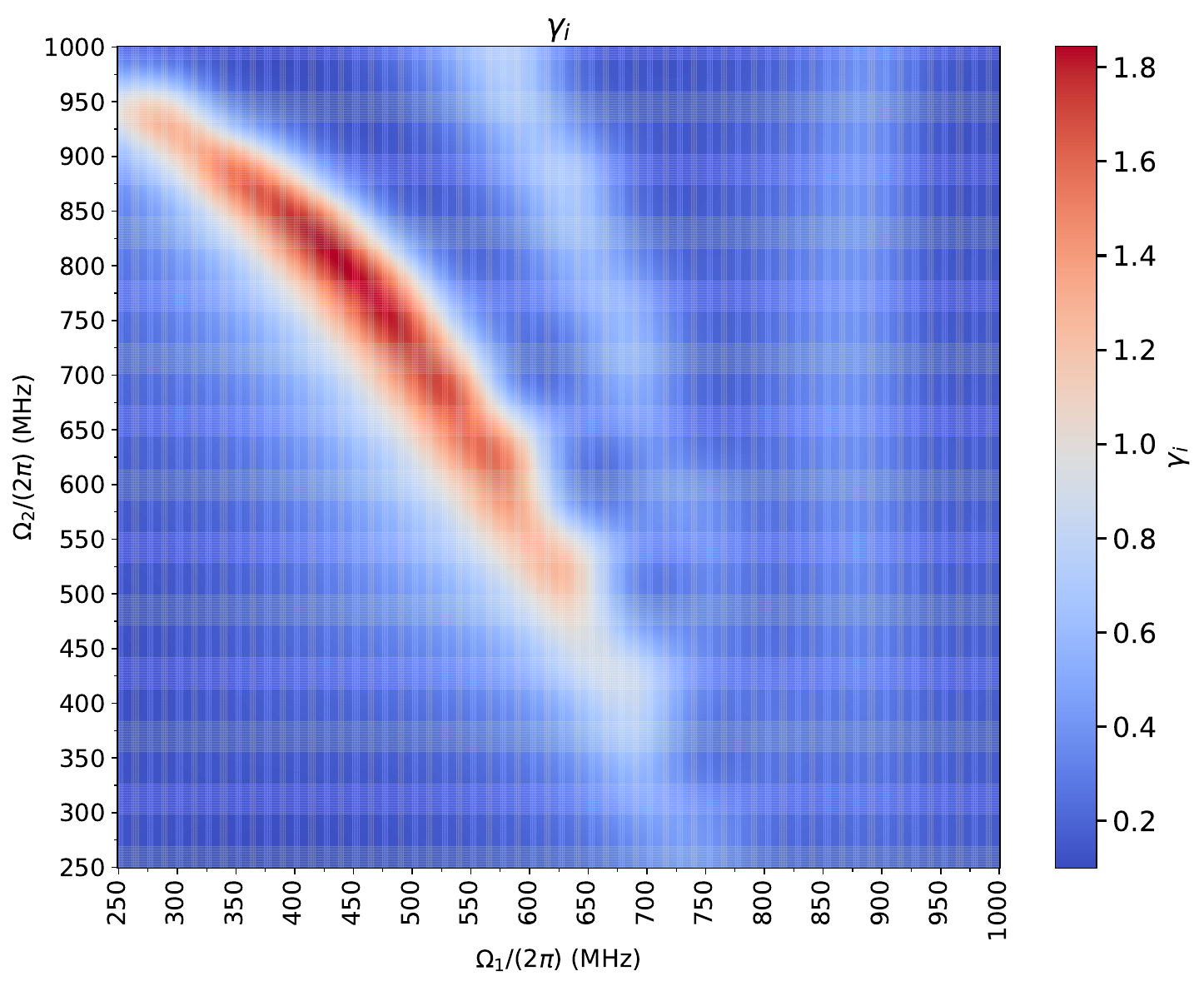} 
   \caption{\parbox{\linewidth}{ 
            \justifying
            Color map in $\Omega_1/(2\pi)$ and $\Omega_2/(2\pi)$ with a step size of 25~MHz for the 2-OS efficiency parameter $\gamma_i$ at an energy $E_{\mathrm{col}} = k_B \times 300$~nK and for $ \Delta/(2\pi)=-25$~MHz.}}
    \label{fig:gamma_color_map}
\end{figure}

The latter quasi-resonant character of 2-OS is best exemplified in Fig. \ref{fig:rates_color_map_D_25_step_50_25} which provides a global view of our results as color maps for $\beta_i^{\text{el}}$, $\beta_i^{\text{inel}}$ and $\beta_i^{\text{rea}}$ as functions of $\Omega_1/(2\pi)$ and $\Omega_2/(2\pi)$ varying from 300~MHz to 1000~MHz, for a fixed detuning $\Delta/(2\pi) = -25$~MHz. The overall patterns of each rate coefficient are similar with three stripes, but are slightly shifted with respect to each other from one panel to the other. This leads to Fig. \ref{fig:gamma_color_map} where only one strip of peak values of $\gamma_i \approx 2$ values  remains in the parameter space ($\Omega_1,\Omega_2$). Again, in the perspective of experimental implementation, we see in Figure \ref{fig:rates_small_step} that the widths in $\Delta$ over which the rate coefficients abruptly vary are quite different for various detuning magnitudes, and not always associated to a quasi-resonant pattern.  

\begin{figure}[t]
    \centering
    \includegraphics[scale=0.32]{rates_380_880_456_831_v2} 
 \caption{\parbox{\linewidth}{ 
            \justifying 
            Collision rate coefficients $\beta_i^{\text{el}}$ (magenta line), $\beta_i^{\text{inel}}$ (blue line), $\beta_i^{\text{rea}}$ (orange line) (all labeled as $\beta$ on the vertical axis, for simplicity) at an energy $E_{\mathrm{col}} = k_B \times 300$~nK, as functions of $\Delta$ for two ($\Omega_1,\Omega_2$) pairs presented in Table \ref{tab:data} calculated with a small step size of 0.05~MHz for $\Delta/(2\pi)$, highlighting resonant features. }}
 \label{fig:rates_small_step}
\end{figure}

\begin{figure}[t]
    \centering    \includegraphics[width=0.96\linewidth]{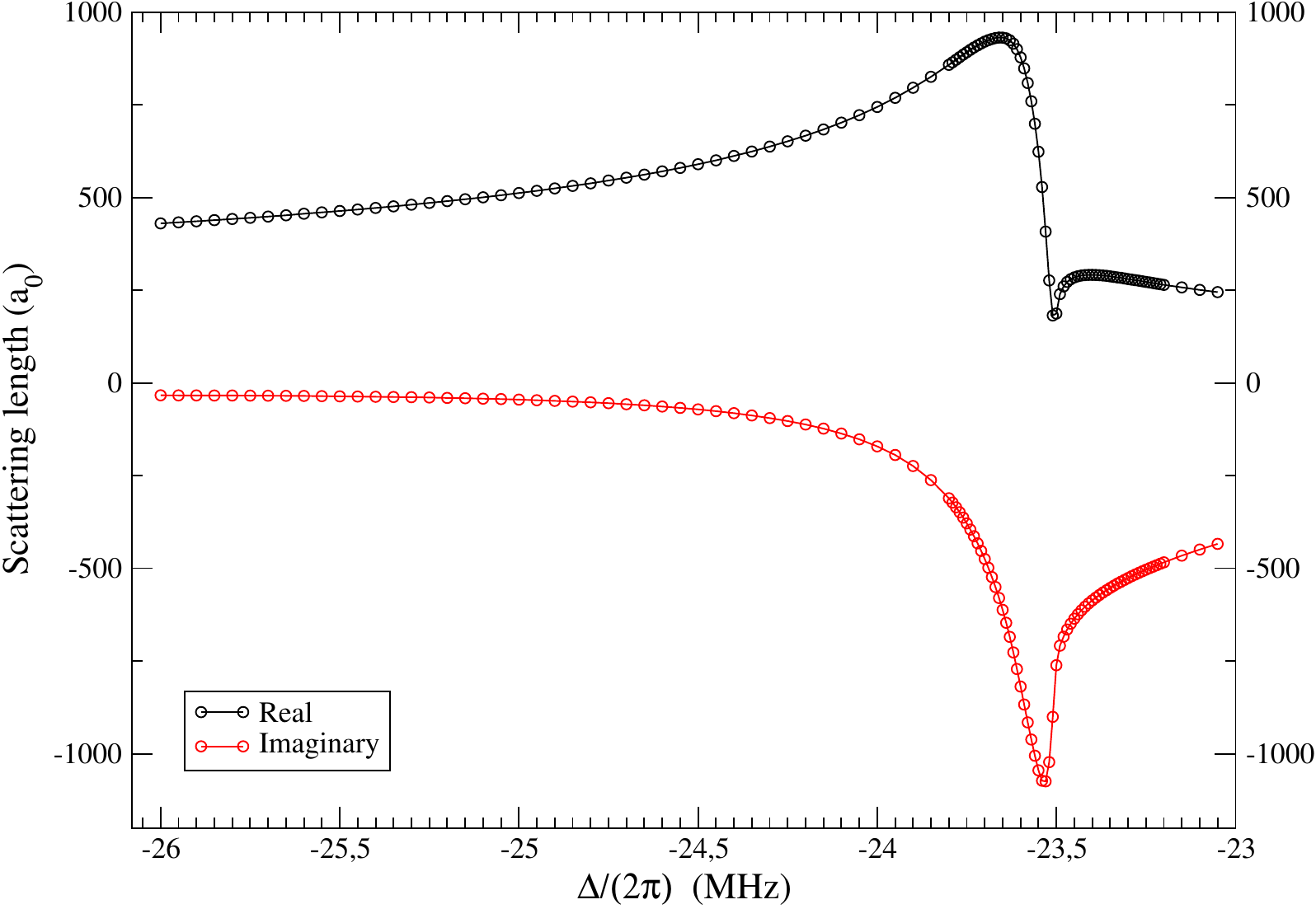}
 \caption{\parbox{\linewidth}{ 
            \justifying
            Real part (black curve) and imaginary part (red curve) of the $S$ matrix element for the elastic collisions expressed as scattering lengths for $\Omega_1/(2\pi)=380$~MHz and $\Omega_2/(2\pi)=880$~MHz as in Fig. \ref{fig:rates_small_step}a. A step of 0.05~MHz is used for $\Delta/(2\pi).$
}}
 \label{fig:scattering-length}
\end{figure}

\begin{figure}[!ht]
    \centering
   \includegraphics[width=0.94\linewidth]{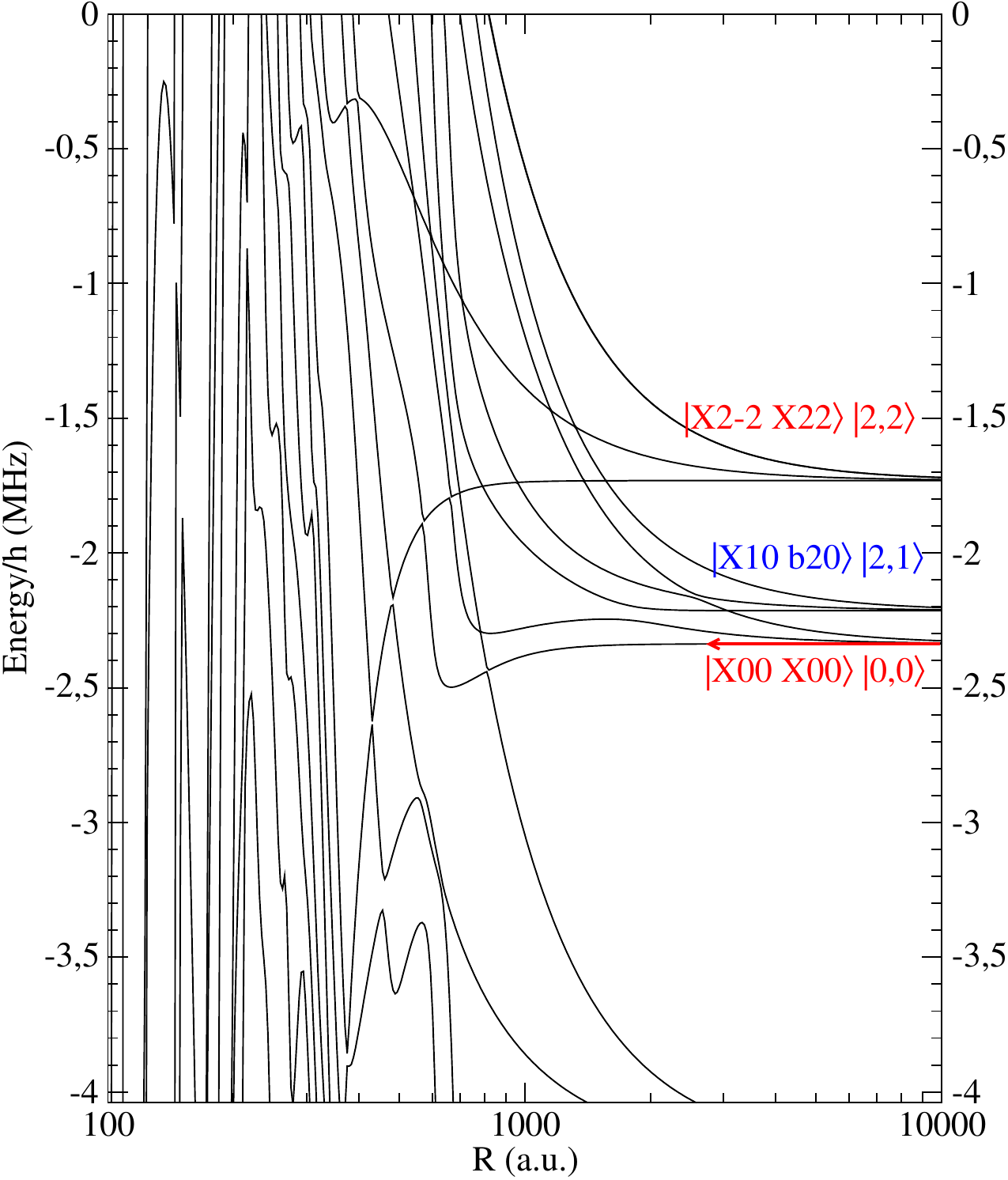}
    \caption{\parbox{\linewidth}{          \justifying
            Same adiabatic dressed PECs as in Fig. \ref{fig:dressed_blue_D_4_O_456_831_ID}b for $\Omega_1/(2\pi)=456$~MHz and $\Omega_2/(2\pi)=831$~MHz but for a blue detuning $\Delta=-2\pi \times 24$~MHz. The curves are labeled as in Fig. \ref{fig:Dressed_pecs_non_eff}. The red arrow shows the entrance channel for the collision. A long-range potential well induced by the lasers in the PEC of the entrance channel is clearly visible, in contrast to Fig. \ref{fig:dressed_blue_D_4_O_456_831_ID}b.}}
    \label{fig:dressed_blue_D_24_O_456_831_ID}
\end{figure}

The sharp structures might be consistent with near-threshold-dressed resonances: as $\Delta$ is varied, a dressed bound (or quasi-bound) state crosses the scattering threshold, strongly modifying coupling to loss channels and producing localized enhancements or suppression of inelastic or reactive collisions. In order to check this hypothesis, we display in Fig. \ref{fig:scattering-length} the variation of the real part and the imaginary part of the elastic $S$ matrix element represented as  scattering lengths of the entrance channel depending on $\Delta$, which clearly exhibits a resonant pattern in the real part, also with a significant imaginary part, as expected from Fig. \ref{fig:rates_small_step}. This description is consistent with the dressed adiabatic PECs of Fig. \ref{fig:dressed_blue_D_24_O_456_831_ID} with $\Omega_1=456$~MHz and $\Omega_2=380$~MHz, where a long-range potential well induced by the presence of the two lasers is now visible for $\Delta/(2\pi)=-24$~MHz, in striking contrast with the pattern of the PECs at $\Delta/(2\pi)=-4$~MHz (Fig. \ref{fig:dressed_blue_D_4_O_456_831_ID}b). A similar pattern has been discussed in several works \cite{avdeenkov2003,lassabliere2018,chen2023,chen2024}. 

\vspace{2cm}

\section{Conclusions}
\label{sec:discussion}

We have presented a theoretical and numerical investigation of two-photon optical shielding (2-OS) for ultracold collisions between dipolar $^{23}$Na$^{39}$K molecules in their rovibronic ground state. Individual molecules are exposed to two lasers tuned to a two-photon Raman resonance arranged as a $\Lambda$-scheme involving an electronically excited molecular state so that photon scattering is suppressed. The colliding molecules are dressed with the laser fields so that their long-range interactions are tunable with respect to the Rabi frequencies and the detunings of the lasers. During collision, the molecular pair is thus described as a dynamical five-level (or double $\Lambda$-) scheme generated by intermolecular interactions. We focused on the regime where detunings are smaller than the Rabi frequencies to engineer the long-range interactions to make them repulsive at large distances and thus suppress short-range dynamics of the molecular pair. 

Our results reveal for the first time the ability of such an arrangement with two optical photons to induce long-range repulsive interactions between laser-dressed ultracold molecules. The efficiency of the resulting shielding is still moderate with respect to the approach based on microwave fields, with rate coefficients for elastic collisions exceeding those for inelastic and reactive collisions by only a factor of 2. The 2-OS conditions are found to be very sensitive to the detuning of the lasers (at the MHz level), while they are less sensitive to the precise choice of Rabi frequencies (at the level of a few tens of MHz), thus representing promising conditions for future experimental implementation. Color maps of the collisional rates confirm the presence of narrow resonance-like strips of favorable conditions in the parameter space of Rabi frequencies. This analysis is confirmed by the presence of resonant structures in the $S$ matrix elements that describe elastic collisions between ground-state molecules, presented as scattering lengths. We show that these features are associated with the appearance of tiny long-range potential wells in the entrance channel, thus enhancing the resonant character of the collisions. 

These findings highlight the potential of 2-OS as a viable and tunable technique for stabilizing ultracold molecular samples, especially in systems where photon scattering or strong microwave fields are undesirable. Although the maximum achieved values of the 2-OS efficiency remain modest compared to those of other techniques, the observed suppression of reactive collisions and the emergence of repulsive long-range laser-induced interactions between molecules suggest a promising path for further optimization. Future work will explore the combination of the present two-photon scheme with static electric fields to induce first-order dipole-dipole interactions, which would then better mimic the approach using microwave fields, while approaching realistic experimental conditions of typical ultracold molecule experiments, where a static electric field is expected to induce anisotropic interactions in forthcoming molecular quantum degenerate gases.

\section*{Acknowledgements}
Stimulating discussions with Prof. Eberhard Tiemann and Prof. Silke Ospelkaus (IQO, Leibniz University, Hannover) are gratefully acknowledged. This work is supported by the grant ANR-22-CE92-0069-01 (OPENMINT project) from the Agence Nationale de la Recherche. L.K. thanks the Deutsche Forschungsgemeinschaft (DFG, German Research Foundation) for support through the Heisenberg Programme—Grant No. 506287139. Calculations have been performed using HPC resources from DNUM CCUB (Centre de Calcul de l’Université de Bourgogne). 

\bibliographystyle{ieeetr}
\bibliography{bibliocold}

@string{as   = "Am. Sci."}

@article{aldegunde2017,
  title = {Hyperfine structure of alkali-metal diatomic molecules},
  author = {Aldegunde, Jesus and Hutson, Jeremy M.},
  journal = {Phys. Rev. A},
  volume = {96},
  issue = {4},
  pages = {042506},
  year = {2017}
}

@article{anderegg2021,
AUTHOR={L. Anderegg and S. Burchesky and Y. Bao and S. S. Yu and T. Karman and E. Chae and K.-K. Ni and W. Ketterle and J. M. Doyle },
TITLE={Observation of microwave shielding of ultracold molecules },
JOURNAL={Science},
YEAR={2021},
VOLUME={373},
NUMBER={},
PAGES={779},
MONTH={},
NOTE={}
}

@article{avdeenkov2003,
AUTHOR={A. V. Avdeenkov and John L. Bohn},
TITLE={Linking Ultracold Polar Molecules},
JOURNAL={Phys. Rev. Lett.},
YEAR={2003},
VOLUME={90},
NUMBER={},
PAGES={043006},
MONTH={},
NOTE={}
}

@article{avdeenkov2006a,
AUTHOR={A. V. Avdeenkov and M. Kajita and J. L. Bohn},
TITLE={Suppression of inelastic collisions of polar [sup 1] Sigma state molecules in an electrostatic field},
JOURNAL={Phys. Rev. A},
YEAR={2006},
VOLUME={73},
NUMBER={},
PAGES={022707},
MONTH={},
NOTE={}
}

@article{bali1994,
  author={S. Bali and D. Hoffmann and T. Walker},
  title={Novel Intensity Dependence of Ultracold Collisions Involving Repulsive States},
  journal={Europhys. Lett.},
  volume={27},
  pages={273},
  year={1994},
}

@article{bause2023,
author = {Bause, Roman and Christianen, Arthur and Schindewolf, Andreas and Bloch, Immanuel and Luo, Xin-Yu},
title = {Ultracold Sticky Collisions: Theoretical and Experimental Status},
journal = {J. Phys. Chem. A},
volume = {127},
number = {3},
pages = {729-741},
year = {2023},
}

@article{bigagli2023,
    author = "{N. Bigagli} and {C. Warner} and {W. Yuan} and {S. Zhang} and {I. Stevenson} and {T. Karman} and {S. Will}",
    title = "Collisionally stable gas of bosonic dipolar ground state molecules",
    year = "2023",
    journal = "Nature Physics",
    volume = "19",
    pages = "1579",
}

@article {bohn2017,
	author = {Bohn, John L. and Rey, Ana Maria and Ye, Jun},
	title = {Cold molecules: Progress in quantum engineering of chemistry and quantum matter},
	journal = {Science},
	volume = {357},
	number = {6355},
	pages = {1002},
	year = {2017}
}

@article{carr2009,
  author={L. D. Carr and J. Ye},
  title={Focus on Cold and Ultracold Molecules},
  journal={New J. Phys.},
  volume={11},
  pages={055009},
  year={2009},
}

@article{karam2025,
  title = {Two-photon-assisted collisions in ultracold gases of polar molecules},
  author = {Karam, Charbel and Hovhannesyan, Gohar and Vexiau, Romain and Lepers, Maxence and Bouloufa-Maafa, Nadia and Dulieu, Olivier},
  journal = {Phys. Rev. A},
  volume = {112},
  issue = {4},
  pages = {043317},
  numpages = {15},
  year = {2025},
  month = {Oct},
  publisher = {American Physical Society},
  doi = {10.1103/9q1s-pjrs},
  url = {https://link.aps.org/doi/10.1103/9q1s-pjrs}
}

@article{chen2023,
  title = {Field-linked resonances of polar molecules},
  author = {Xing-Yan Chen and Andreas Schindewolf and Sebastian Eppelt and Roman Bause and Marcel Duda and Shrestha Biswas and Tijs Karman and Timon Hilker and Immanuel Bloch and Xin-Yu Luo},
  journal = {Nature},
  volume = {614},
  pages = {59-63},
  year = {2023},
  doi = {https://doi.org/10.1038/s41586-022-05651-8}
}

@article{chen2024,
  title={Ultracold field-linked tetratomic molecules},
  author={Chen, Xing-Yan and Biswas, Shrestha and Eppelt, Sebastian and Schindewolf, Andreas and Deng, Fulin and Shi, Tao and Yi, Su and Hilker, Timon A and Bloch, Immanuel and Luo, Xin-Yu},
  journal={Nature},
  volume={626},
  number={7998},
  pages={283--287},
  year={2024},
  publisher={Nature Publishing Group UK London}
}

@article{cornish2024,
  title={Quantum computation and quantum simulation with ultracold molecules},
  author={Cornish, Simon L and Tarbutt, Michael R and Hazzard, Kaden RA},
  journal={Nature Physics},
  volume={20},
  number={5},
  pages={730--740},
  year={2024},
  publisher={Nature Publishing Group UK London}
}

@article{demille2024,
  title={Quantum sensing and metrology for fundamental physics with molecules},
  author={DeMille, David and Hutzler, Nicholas R and Rey, Ana Maria and Zelevinsky, Tanya},
  journal={Nature Physics},
  volume={20},
  number={5},
  pages={741--749},
  year={2024},
  publisher={Nature Publishing Group UK London}
}

@article{doyle2004,
AUTHOR={J. Doyle and  B. Friedrich and  R. Krems  and F. Masnou-Seeuws},
TITLE={Quo vadis, cold molecules},
JOURNAL={Eur. Phys. J. D},
YEAR={2004},
VOLUME={31},
NUMBER={},
PAGES={149-164},
MONTH={december},
NOTE={}
}

@article{dulieu2009,
AUTHOR={O. Dulieu and  C. Gabbanini},
TITLE={Formation and interactions of cold and ultracold molecules : new challenges for interdisciplinary physics},
JOURNAL={Rep. Prog. Phys.},
YEAR={2009},
VOLUME={72},
NUMBER={},
PAGES={086401},
MONTH={},
NOTE={}
}

@article{gonzalez-martinez2017,
  title = {Adimensional theory of shielding in ultracold collisions of dipolar rotors},
  author = {Gonz\'alez-Mart\'{\i}nez, Maykel L. and Bohn, John L. and Qu\'em\'ener, Goulven},
  journal = {Phys. Rev. A},
  volume = {96},
  pages = {032718},
  year = {2017}
}

@article{gregory2019,
  author={P. D. Gregory and M. D. Frye and J. A. Blackmore and E. M. Bridge and R. Sawant and J. M. Hutson and and S. L. Cornish},
  title={Sticky collisions of ultracold RbCs molecules},
  journal={Nature Comm.},
  volume={10},
  pages={3104},
  year={2019},
}

@article{haroche2025,
  title = {Laser, Offspring and Powerful Enabler of Quantum Science},
  author = {Haroche, Serge},
  journal = {PRX Quantum},
  volume = {6},
  issue = {1},
  pages = {010102},
  numpages = {27},
  year = {2025},
  month = {Mar},
  publisher = {American Physical Society},
  doi = {10.1103/PRXQuantum.6.010102},
  url = {https://link.aps.org/doi/10.1103/PRXQuantum.6.010102}
}

@article{hoffmann1996,
  title = {Trap-depth measurements using ultracold collisions},
  author = {Hoffmann, D. and Bali, S. and Walker, T.},
  journal = {Phys. Rev. A},
  volume = {54},
  pages = {R1030},
  year = {1996},
}

@article{karam2023,
  title = {Two-photon optical shielding of collisions between ultracold polar molecules},
  author = {Karam, Charbel and Vexiau, Romain and Bouloufa-Maafa, Nadia and Dulieu, Olivier and Lepers, Maxence and Borgloh, Mara Meyer zum Alten and Ospelkaus, Silke and Karpa, Leon},
  journal = {Phys. Rev. Res.},
  volume = {5},
  issue = {3},
  pages = {033074},
  numpages = {9},
  year = {2023},
  month = {Aug},
  publisher = {American Physical Society},
  doi = {10.1103/PhysRevResearch.5.033074},
  url = {https://link.aps.org/doi/10.1103/PhysRevResearch.5.033074}
}

@phdthesis{karam2024,
  TITLE = {{Optical shielding of collisions between ultracold polar molecules}},
  AUTHOR = {Karam, Charbel},
  URL = {https://theses.hal.science/tel-04920980},
  NUMBER = {2024UPASP137},
  SCHOOL = {{Universit{\'e} Paris-Saclay}},
  YEAR = {2024},
  MONTH = Nov,
  TYPE = {Theses},
  HAL_ID = {tel-04920980},
  HAL_VERSION = {v1},
}

@article{karman2018,
  title = {Microwave Shielding of Ultracold Polar Molecules},
  author = {Karman, Tijs and Hutson, Jeremy M.},
  journal = {Phys. Rev. Lett.},
  volume = {121},
  issue = {16},
  pages = {163401},
  numpages = {5},
  year = {2018},
}

@article{karman2024,
  title={Ultracold chemistry as a testbed for few-body physics},
  author={Karman, Tijs and Tomza, Micha{\l} and P{\'e}rez-R{\'\i}os, Jes{\'u}s},
  journal={Nature Physics},
  volume={20},
  number={5},
  pages={722--729},
  year={2024},
  publisher={Nature Publishing Group UK London}
}

@article{kokoouline1999,
AUTHOR={V. Kokoouline and O. Dulieu and R. Kosloff and F. Masnou-Seeuws},
TITLE={Mapped {F}ourier methods for long-range molecules: application to perturbations in the {R}b$_2(0_u^+)$ photoassociation spectrum},
JOURNAL={J. Chem. Phys.},
YEAR={1999},
VOLUME={110},
NUMBER={},
PAGES={9865--9877},
MONTH={},
NOTE={}
}

@article{lassabliere2018,
  title = {Controlling the Scattering Length of Ultracold Dipolar Molecules},
  author = {Lassabli\`ere, Lucas and Qu\'em\'ener, Goulven},
  journal = {Phys. Rev. Lett.},
  volume = {121},
  issue = {16},
  pages = {163402},
  numpages = {6},
  year = {2018},
}

@article{li2011b,
  title = {Electromagnetically induced transparency in an open multilevel system},
  author = {Li, Tian and Lu, Mei-Ju and Weinstein, Jonathan D.},
  journal = {Phys. Rev. A},
  volume = {84},
  issue = {2},
  pages = {023801},
  numpages = {6},
  year = {2011},
  month = {Aug},
  publisher = {American Physical Society},
  doi = {10.1103/PhysRevA.84.023801},
  url = {https://link.aps.org/doi/10.1103/PhysRevA.84.023801}
}

@article{li2021,
  title = {Tuning of dipolar interactions and evaporative cooling in a three-dimensional molecular quantum gas},
  author = {J.-R. Li and W. G. Tobias and K. Matsuda and C. Miller and G. Valtolina and L. De Marco and R. R. Wang and L. Lassabli\`ere and G. Qu\'em\'ener and J. L. Bohn and J. Ye},
  journal = {Nature Phys.},
  volume = {17},
  pages = {1144},
  year = {2021},
}

@article{liu2022bimolecular,
  title={Bimolecular chemistry in the ultracold regime},
  author={Liu, Yu and Ni, Kang-Kuen},
  journal={Annual Rev.Phys. Chem.},
  volume={73},
  number={1},
  pages={73--96},
  year={2022},
  publisher={Annual Reviews}
}

@article{manolopoulos1986,
author = {Alexander,Millard H.  and Manolopoulos,David E. },
title = {A stable linear reference potential algorithm for solution of the quantum close‐coupled equations in molecular scattering theory},
journal =  {J. Chem. Phys.},
volume = {86},
number = {4},
pages = {2044-2050},
year = {1987},

}

@article{marcassa1994,
AUTHOR={L. Marcassa and S. Muniz and E. de Queiroz and S. Zilio and V. Bagnato and J. Weiner and P. S. Julienne and K. A. Suominen},
TITLE={Optical suppression of photoassociative ionization in a magneto-optical trap},
JOURNAL={Phys. Rev. Lett.},
YEAR={1994},
VOLUME={73},
NUMBER={},
PAGES={1911},
MONTH={},
NOTE={}
}

@article{mayle2013,
AUTHOR={M. Mayle and G. Qu\'em\'ener and B. P. Ruzic and J. L. Bohn},
TITLE={Scattering of ultracold molecules in the highly resonant regime},
JOURNAL={Phys. Rev. A},
YEAR={2013},
VOLUME={87},
NUMBER={},
PAGES={012709},
MONTH={},
NOTE={}
}

@article{molony2014,
  title = {Creation of Ultracold $^{87}\mathrm{Rb}^{133}\mathrm{Cs}$ Molecules in the Rovibrational Ground State},
  author = {Molony, P. K. and Gregory, P. D. and Ji, Z. and Lu, B. and K\"oppinger, M. P. and Le Sueur, C. R. and Blackley, C. L. and Hutson, J. M. and Cornish, S. L.},
  journal = {Phys. Rev. Lett.},
  volume = {113},
  pages = {255301},
  year = {2014},
}

@article{moses2017,
author = {S. A. Moses and J. P. Covey and M. T. Miecnikowski and D. S. Jin and J. Ye},
title = {New frontiers for quantum gases of polar molecules},
year = {2017},
journal = {Nature Phys.},
volume = {13},
pages = {13},
}

@article{mukherjee2023,
  title={Shielding collisions of ultracold CaF molecules with static electric fields},
  author={Mukherjee, Bijit and Frye, Matthew D and Le Sueur, C Ruth and Tarbutt, Michael R and Hutson, Jeremy M},
  journal={Phys. Rev. Res.},
  volume={5},
  number={3},
  pages={033097},
  year={2023},
  publisher={APS}
}

@article{ospelkaus2010a,
AUTHOR={S. Ospelkaus and  K.-K. Ni and  D. Wang and  M. H. G. de Miranda and  B. Neyenhuis and G. Qu\'em\'ener and  P. S. Julienne and J. Bohn and D. S. Jin and  J. Ye},
TITLE={Quantum state controlled chemical reactions of ultracold potassium-rubidium molecules},
JOURNAL={Science},
YEAR={2010},
VOLUME={327},
NUMBER={},
PAGES={853},
MONTH={},
NOTE={}
}

@article{park2015,
  title = {Ultracold Dipolar Gas of Fermionic $^{23}\mathrm{Na}^{40}\mathrm{K}$ Molecules in Their Absolute Ground State},
  author = {Park, Jee Woo and Will, Sebastian A. and Zwierlein, Martin W.},
  journal = {Phys. Rev. Lett.},
  volume = {114},
  pages = {205302},
  year = {2015},
}

@article{quemener2012,
  title={Ultracold molecules under control},
  author={G. Qu{\'e}m{\'e}ner and P. S. Julienne},
  journal={Chem. Rev.},
  volume={112},
  pages={4949},
  year={2012},
}

@article{schindewolf2022,
  author = {Schindewolf, Andreas and Bause, Roman and Chen, Xing-Yan and Duda, Marcel and Karman, Tijs and Bloch, Immanuel and Luo, Xin-Yu},  
  title = {Evaporation of microwave-shielded polar molecules to quantum degeneracy},
journal = {Nature},
volume = {607},
pages = {677},
year = {2022},
}

@article{shi2025bose,
  title={Bose-Einstein condensate of ultracold sodium-rubidium molecules with tunable dipolar interactions},
  author={Shi, Zhaopeng and Huang, Zerong and Deng, Fulin and Jin, Wei-Jian and Yi, Su and Shi, Tao and Wang, Dajun},
  journal={arXiv preprint arXiv:2508.20518},
  year={2025}
}

@article{suominen1996a,
  title = {Ultracold collisions and optical shielding in metastable xenon},
  author = {Suominen, K.-A. and Burnett, K. and Julienne, P. S. and Walhout, M. and Sterr, U. and Orzel, C. and Hoogerland, M. and Rolston, S. L.},
  journal = {Phys. Rev. A},
  volume = {53},
  pages = {1678},
  year = {1996},
}

@article{takekoshi2014,
  Author 	= {T. Takekoshi and L. Reichs\"ollner and A. Schindewolf and J. M. Hutson and C. R. Le Sueur and O. Dulieu and F. Ferlaino and R. Grimm and H.-C. N\"agerl},
Title = {Ultracold dense samples of dipolar RbCs molecules in the rovibrational and hyperfine ground state},
  Journal = {Phys. Rev. Lett.},
  Year = {2014},
  Volume = {113},
  pages = {205301}
}

@article{wang2005,
AUTHOR={D. Wang and  E. E. Eyler and  P. L. Gould and W. C. Stwalley},
TITLE={State-selective detection of near-dissociation ultracold KRb X 1 Sigma+ and a 3 Sigma + molecules},
JOURNAL={Phys. Rev. A},
YEAR={2005},
VOLUME={72},
NUMBER={},
PAGES={032502},
MONTH={},
NOTE={}
}

@article{wang2015,
	year = 2015,
	volume = {17},
	number = {3},
	pages = {035015},
	author = {Gaoren Wang and Goulven Qu{\'{e}}m{\'{e}}ner},
	title = {Tuning ultracold collisions of excited rotational dipolar molecules},
	journal = {New J. Phys.}
}

@article{willner2004,
AUTHOR={K. Willner and O. Dulieu and F. Masnou-Seeuws},
TITLE={Mapped grid methods for long-range molecules and cold collisions },
JOURNAL={J. Chem. Phys.},
YEAR={2004},
VOLUME={120},
NUMBER={2},
PAGES={548-561},
MONTH={january},
NOTE={}
}

@article{xie2020,
AUTHOR={T. Xie and  M. Lepers and  R. Vexiau and  A. Orb\'{a}n and  O. Dulieu and N. Bouloufa-Maafa },
TITLE={Optical shielding of destructive chemical reactions between ultracold ground-state NaRb molecules},
JOURNAL={Phys. Rev. Lett.},
YEAR={2020},
VOLUME={125},
NUMBER={},
PAGES={153202},
MONTH={},
NOTE={}
}

@article{xie2022,
AUTHOR={T. Xie and  A. Orb\'{a}n and X. Xing and  E. Luc-Koenig and  R. Vexiau  and  O. Dulieu and N. Bouloufa-Maafa },
TITLE={Engineering long-range interactions between ultracold atoms with light},
JOURNAL={J. Phys. B: At. Mol. Opt. Phys.},
YEAR={2022},
VOLUME={ 55},
NUMBER={},
PAGES={034001},
MONTH={},
NOTE={}
}
\end{document}